\def\papersize{letterpaper}
\documentclass[11pt,\papersize]{article}
\usepackage{graphicx}
\usepackage[subrefformat=parens,labelformat=parens]{subfig}

\usepackage{url}

\usepackage{amsmath}
\usepackage{algorithmic}
\usepackage{float}
\usepackage{authblk}
\usepackage{amssymb, amsthm, mathrsfs, amsopn}
\usepackage{threeparttable}
\usepackage{booktabs}
\usepackage{multirow}
\usepackage{adjustbox}
\usepackage{verbatim}
\usepackage{algorithm}
\usepackage{enumerate}
\usepackage[shortlabels]{enumitem}
\newlist{steps}{enumerate}{1}
\setlist[steps, 1]{label = Step \arabic*:}
\usepackage{xcolor}

\usepackage[square,sort&compress,comma,numbers]{natbib}

\theoremstyle{remark}

\theoremstyle{definition}

\usepackage{setspace}
\usepackage[\papersize,top=1in, bottom=1in, left=1in, right=1in]{geometry}

\title{\vskip -0.8in An Adapted Geographically Weighted Lasso (Ada-GWL) model for estimating metro ridership}

\author[1]{Yuxin He\thanks{email: yuxinhe2-c@my.cityu.edu.hk}}
\author[2]{Yang Zhao\thanks{Corresponding author, email: yang.zhao@my.cityu.edu.hk}}
\author[1,2]{Kwok Leung Tsui\thanks{email: kltsui@cityu.edu.hk}}
\affil[1]{School of Data Science\\
City University of Hong Kong\\
Kowloon, Hong Kong}
\affil[2]{Centre for Systems Informatics Engineering\\
City University of Hong Kong\\
Kowloon, Hong Kong}

\providecommand{\keywords}[1]{\textbf{\textit{Keywords---}} #1}

\makeatletter
\newenvironment{breakablealgorithm}
{% \begin{breakablealgorithm}
	\begin{center}
		\refstepcounter{algorithm}% New algorithm
		\hrule height.8pt depth0pt \kern2pt% \@fs@pre for \@fs@ruled
		\renewcommand{\caption}[2][\relax]{% Make a new \caption
			{\raggedright\textbf{\ALG@name~\thealgorithm} ##2\par}%
			\ifx\relax##1\relax % #1 is \relax
			\addcontentsline{loa}{algorithm}{\protect\numberline{\thealgorithm}##2}%
			\else % #1 is not \relax
			\addcontentsline{loa}{algorithm}{\protect\numberline{\thealgorithm}##1}%
			\fi
			\kern2pt\hrule\kern2pt
		}
	}{% \end{breakablealgorithm}
		\kern2pt\hrule\relax% \@fs@post for \@fs@ruled
	\end{center}
}
\makeatother
\begin{document}
\maketitle
\begin{abstract}
Ridership estimation at station level plays a critical role in metro transportation planning. Among various existing ridership estimation methods, direct demand model has been recognized as an effective approach. However, existing direct demand models including Geographically Weighted Regression (GWR) have rarely included local model selection in ridership estimation. In practice, acquiring insights into metro ridership under multiple influencing factors from a local perspective is important for passenger volume management and transportation planning operations adapting to local conditions. In this study, we propose an Adapted Geographically Weighted Lasso (Ada-GWL) framework for modelling metro ridership, which performs regression-coefficient shrinkage and local model selection. It takes metro network connection intermedia into account and adopts network-based distance metric instead of Euclidean-based distance metric, making it so-called adapted to the context of metro networks. The real-world case of Shenzhen Metro is used to validate the superiority of our proposed model. The results show that the Ada-GWL model performs the best compared with the global model (Ordinary Least Square (OLS), GWR, GWR calibrated with network-based distance metric and GWL in terms of estimation error of the dependent variable and goodness-of-fit.  Through understanding the variation of each coefficient across space (elasticities) and variables selection of each station, it provides more realistic conclusions based on local analysis. Besides, through clustering analysis of the stations according to the regression coefficients, clusters' functional characteristics are found to be in compliance with the facts of the functional land use policy of Shenzhen. These results of the proposed Ada-GWL model demonstrate a great spatial explanatory power in transportation planning.
\end{abstract}
\section{Introduction}\label{s.intro}

With the rapid urbanization, metro has become the mainstream of public transportation for its advantages such as fast speed, large capacity, and convenience. At the stage of urban construction and transportation planning, it is crucial to explore multiple urban indicators from a systematic perspective to capture the urban traffic-related phenomenon like metro ridership at station level, which is a critical element for determining the scale of stations and access facilities. Metro ridership at station level is known to be influenced by the interactions of the different components of an urban system (e.g., land-use, socio-economics, etc.). Understanding the influence of these components is vital to accurately estimate travel demand and to effectively make design schemes of urban systems including the identification of which public infrastructures, services, and resources need to be built and deployed. Modelling metro ridership at station level can help to not only estimate and forecast ridership but also analyse the influencing factors on it. 

Various methods have been proposed for transit ridership estimation. As one of the best-known models, the four-step (generation, distribution, mode choice, and assignment) model has dominated the history of transport modeling since the 1950s \citep{mcnally2000four}.  However, the four-step model has many drawbacks in practice \citep{marshall2006sketch}, such as limitation in model accuracy, low data precision, insensitivity to land use, institutional barriers, and high expense \citep{gutierrez2011transit}. The four-step model is generally effective for estimating transit ridership on a regional scale rather than more detailed scales (such as station level) \citep{cardozo2012application}. As an alternative to the four-step model, direct demand models have drawn growing attention for ridership estimation in recent decades. Direct demand models estimate ridership as a function of influencing factors within the Pedestrian Catchment Areas (PCA) via regression analysis, which enable identifying factors that contribute to higher transit ridership\citep{gutierrez2011transit,choi2012analysis,cervero2006alternative,kuby2004factors,chu2004ridership}. In the models, a PCA is a geographic area for which a station attracts passengers. The size and shape of a catchment area depend on how accessible a station is and how far it is from alternative stations. One can use buffers to create circular catchment areas by a specific distance or use Thiessen polygons to illustrate the area most accessible to each station. The major advantages of direct demand models in travel analysis are simplicity of use, easy interpretation of results, immediate response, and low cost. The direct demand models generally use OLS multiple regression , which assumes parametric stability. With the development of spatial modeling, direct demand models could increase their spatial explanatory power by using GWR, which is designed to model spatial parametric nonstationarity, and variance heterogeneity. However, the issues of GWR models, such as collinearity in the estimated coefficients, and network connection intermedia may influence metro ridership modelling and estimation. Thus, there are still a number of limitations to be promoted in the existing research on transit ridership modeling.

In light of deficiencies of current direct demand models, considering the pros and cons of Geographically Weighted Regression (GWR) for modelling potentially spatially varying relationships,  this paper proposes an Adapted Geographically Weighted Lasso (Ada-GWL) framework for modelling the ridership of metro systems, which considers using network-based distance metric rather than Euclidean-based distance metric, and simultaneously shrink regression coefficients and select feature locally in the process of GWR with Least Absolute Shrinkage and Selection Operator (LASSO).The first objective of this study is to identify the association between multiple factors and metro station ridership from a local perspective. The second objective is to estimate ridership at station level accurately. An Ada-GWL model is used to analyze the significance of various local factors that impact metro station ridership. The ridership data of Shenzhen Metro at station-level in the year of 2013 is used to validate the superiority of our proposed model over global model (Ordinary Least Square), GWR, GWR calibrated with network-based distance metric, and GWL model. 

Our contributions are three-fold:

(1) This study introduces network-based distance metric during estimating  ridership. By implementing the Ada-GWL model on the real world case study of Shenzhen metro systems, we calibrate distance with network-based distance metric instead of Euclidean-based distance metric. The reason is that locations may be related through a certain intermedia system but not in Euclidean space. This paper investigates ridership at metro stations, which are related through railways. In practice,  taking network connection intermedia into consideration by using network-based distance metric is important as routine transportation activities normally occur on a determined transportation network but not a continuous space \citep{zhang2014transit}. Nonetheless, few existing direct demand models have included spatial aspects considering network connection intermedia.

(2) We adopt the basic framework of GWR considering spatial autocorrelation of variables to model metro ridership, which outperforms the traditional global regression model OLS in terms of model fitting and spatial explanatory power.

(3) This study enables simultaneous coefficient penalization and model selection during GWR. Even though GWR models have been utilized to model and analyze the transit ridership and its influencing factors, the collinearity of GWR may make interpreting individual coefficients problematic. Also, different  influencing factors may exist at different stations, so the feature selection of the model can be conducted from the local instead of global perspective, that is, local penalization and feature selection, which have rarely been done in GWR research.

In general, to the extent of our knowledge, this is the first work that investigates the utility of GWL model calibrated with network-based distance metric for estimating metro ridership and analyzing the influencing factors from a local perspective. It is worth noting that the fundamental framework proposed is not limited to the specific dataset used in this study, but can be extended and generalized to other applications in transportation systems, such as light rail and rapid bus transit systems.

The remainder of this paper is organized as follows. In Section 2, we review the current literature on transit ridership estimation modeling and factors investigated in those models; In Section 3, we present the data description and various influencing factors considered; In Section 4, we describe the methodology for estimating metro transit ridership and identifying significant influencing factors. Section 5 provides the details for model implementation and results analysis and discussion. Finally, we conclude this paper in Section 6.

\section{Literature Review}
\subsection{Transit Ridership Estimation Modelling}
This paper summarizes the related studies on direct demand models for ridership estimation (see Figure \ref{Fig1} in \ref{a.litertab}). As summarized in Figure \ref{Fig1}, the most widely used method is Ordinary Least Squares (OLS) or linear least squares in the study of the influencing factors on transit ridership by using direct demand models, Kuby et al.(2004), Sohn and Shim(2010), Loo et al.(2010), Sung and Oh(2011), Gutierrez et al.( 2011), Thompson et al.(2012), Guerra et al.(2012), Zhao et al.(2013), Chan and Miranda-Moreno (2013), Singhai et al.(2014), Liu et al.(2016) and Pan et al.(2017) all applied OLS regression to model transit ridership and its influencing factors\citep{kuby2004factors,sohn2010factors,loo2010rail,sung2011transit,gutierrez2011transit,thompson2012really,guerra2012half,zhao2013influences,chan2013station,singhal2014impact,liu2016increase,pan2017determines}. However, the limitation of OLS models is that they assume the transit ridership is affected by various factors but has nothing to do with the spatial location, without considering the spatial autocorrelation, in other words, OLS models estimate ridership from a global perspective believing that calculated coefficients do not have significant differences in space. Fotheringham (1996)  proposed Geographically Weighted Regression (GWR) model which can reveal the spatial relations on the condition of spatial heterogeneity. GWR model has a strong capability of spatial data analysis \citep*{brunsdon1996geographically}, and it has been widely used to analyse spatial data in economics, geography, ecology, and many other fields of research. At present, Cardozo et al.(2012) compared the performance of OLS and GWR in modelling transit ridership and its influencing factors, and GWR showed the better goodness-of-fit than OLS for forecasting station-level ridership\citep{cardozo2012application}. In addition, Jun et al.(2015) used mixed geographically weighted regression (MGWR) models consisting of local and global independent variables to examine the relationship between the land use characteristics and subway ridership\citep{jun2015land}. However, there was still room for improvement in their GWR models, such as considering the impact of network connection intermedia on distance metric and correlation in the estimated coefficients during GWR \citep{wheeler2009simultaneous}. Actually, Wheeler and Tiefelsdorf (2005) showed that GWR coefficients can be correlated even though there is no independent variable collinearity, the coefficient correlation increases systematically with increasing collinearity\citep*{wheeler2005multicollinearity}. Therefore, the application of GWR in the estimation of transit ridership needs to be promoted.

The research of GWR with penalized forms (e.g.,  $\ell 1$ norm/  $\ell 2$ norm) can be found in Wheeler's studies\citep{wheeler2007diagnostic,wheeler2009simultaneous}. Both ridge regression ($\ell 2$ norm) and the Lasso ($\ell 1$ norm) are penalization methods which impose constraints on the regression coefficients. Ridge regression aims to alleviate collinearity effects by making use of the $\ell 2$ penalty to regularize the size of regression coefficients and reducing variance. The Lasso also performs regularization on variables in consideration. It considers the absolute values of the sum of the regression coefficients. It also sets the coefficients to zero thus simultaneously performing regression coefficient penalization and model selection. Wheeler (2007) implemented a ridge-regression version of GWR, called GWRR, and found it could constrain the regression coefficients to reduce local correlation effects in an existing dataset and showed a lower estimation error for the dependent variable than of GWR \citep{wheeler2007diagnostic}. The lasso has also been introduced into the GWR framework in Wheeler (2009)'s study. The lasso is considered in the GWR framework
since it has capability to simultaneously coefficient regularization and local model selection, also for its potential to reduce prediction and estimation errors for estimating the dependent variable in GWR. Wheeler (2009) implemented a Lasso version of GWR, called Geographically Weighted Lasso (GWL), and demonstrated the benefit of GWL through a comparative analysis with GWR and GWRR.It can stabilize regression coefficients in the presence of collinearity and produce lower estimation error of the dependent variable than does GWR, and also offered a key advantage to GWRR for the identification of insignificant local effects by doing model selection\citep{wheeler2009simultaneous}. However, the existing version GWL model didn't consider the effect of network connection intermedia, which opened up new perspectives that fully justified the pursuit of this research. Thus, we try to design a version of GWL model adapted to metro network that considering network connection intermedia during calculating distance.

A comprehensive review of direct demand models can also be found in the work by Walters and Cervero (2003) and Cardozo et al. (2012) \citep{walters2003forecasting,cardozo2012application}. 

\subsection{Dependent Variables and Independent Variables}
With regard to the dependent variables of direct demand models considered, daily ridership was selected in most of the relevant studies (also see Figure \ref{Fig1} in \ref{a.litertab}), such as the research of Kuby et al.(2004), Chu (2004), Sohn and Shim (2010) , Loo et al.(2010), Zhao et al.(2013) and Zhang and Wang(2014), which all took the average weekday ridership as dependent variable\citep{kuby2004factors,chu2004ridership,sohn2010factors,loo2010rail,zhao2013influences,zhang2014transit}. In addition, monthly station ridership was considered in the research of Gutierrez et al.(2011)\citep{gutierrez2011transit}. For shorter time span, Zhao et al.(2013), Singhal et al.(2014), Hu et al.(2016), Li et.al (2016) and Chan and Miranda-Moreno (2013)chose the hourly (eg. Peak hour) ridership as the dependent variable of their models\citep{zhao2013influences,singhal2014impact,hu2016impacts,li2016forecasting,chan2013station}.  

Concerning independent variables of models, they can be roughly divided into the following categories, Land use variables, social economic variables, accessibility and the network structure. Figure \ref{Fig1} summarized the independent variables considered in existing studies. For the first kind of variables, such as the research of Estupinan and Rodriguez (2008), Loo et al. (2010), Sohn and Shim (2010), Gutierrez et al. (2011), Sung and Oh (2011), Choi et al. (2012), Cardozo et al. (2012) and Zhao et al (2013)\citep{estupinan2008relationship,loo2010rail,sohn2010factors,gutierrez2011transit,sung2011transit,choi2012analysis,cardozo2012application,zhao2013influences}, which considered commercial, residence, education, entertainment and other mixed land use as independent variables. For the second kind of variables, factors considered mainly are population, employment, and automobile ownership ratio. For example, Chu (2004), Kuby et al. (2004), Loo et al. (2010), Sohn and Shim (2010), Gutierrez et al. (2011), Choi et al. (2012), Cardozo et al. (2012), Guerra et al. (2012), Thompson et al. (2012) and Zhao et al. (2013) analyzed the influence of population and employment on transit ridership \citep{chu2004ridership,kuby2004factors,loo2010rail,sohn2010factors,gutierrez2011transit,choi2012analysis,cardozo2012application,guerra2012half,thompson2012really,zhao2013influences}. Chu (2004), Loo et al. (2010), Thompson et al (2012), Cardozo et al. (2012) and Zhao et al. (2013) considered the relationship between automobile ownership and the ridership\citep{chu2004ridership,loo2010rail,thompson2012really,cardozo2012application,zhao2013influences}. For the third category, accessibility, Chu (2004), Kuby et al. (2004), Estupinan and Rodriguez (2008), Loo et al. (2010), Sung and Oh (2011), Gutierrez et al. (2011), Choi et al. (2012), Cardozo et al. (2012), Guerra et al. (2012) and Zhao et al. (2013) analyzed the influence of bus feeder system on transit ridership\citep{chu2004ridership,kuby2004factors,estupinan2008relationship,loo2010rail,sohn2010factors,thompson2012really,choi2012analysis,guerra2012half,zhao2013influences}. Moreover, Kuby et al. (2004), Estupinan and Rodriguez (2008), Loo et al. (2010), Sohn and Shim (2010), Thompson et al. (2012), Choi et al. (2012), Guerra et al. (2012) and Zhao et al. (2013) studied on the station accessibility\citep{kuby2004factors,estupinan2008relationship,loo2010rail,sohn2010factors,thompson2012really,choi2012analysis,guerra2012half,zhao2013influences}. Finally, with regard to network structure, Kuby et al. (2004) took the influence of transfer station and terminal station on ridership into account, but both of them were regarded as dummy variables in their models\citep{kuby2004factors}. So far, few relevant studies have been carried out from the quantitative perspective by using measurements from the field of complex network, such as quantifying the practical significance (e.g., transfer, terminal station and the importance of stations' positions that lie on a large number of paths taken by flows) by the calculation of nodes' degree centrality and betweenness centrality in networks.

\section{Methodology}
\subsection{Geographically Weighted Regression}
GWR model is an extension of OLS, which can be formulated as:
\begin{equation} 
{y}_{i}=\beta +\underset{k=1}{\overset{p}{\mathop \sum }}\,{{\beta }_{k}}{{x}_{ik}}+{{\varepsilon }_{i}}   \label{e.ols}
\end{equation}

When geographical location factors are introduced into regression parameters to allow local parameter estimation, OLS model evolves into GWR model, which is as following:
\begin{equation} 
{y}_{i}={{\beta }_{0}}\left( {{u}_{i}},{{v}_{i}} \right)+\underset{k=1}{\overset{p}{\mathop \sum }}\,{{\beta }_{k}}\left( {{u}_{i}},{{v}_{i}} \right){{x}_{ik}}+{{\varepsilon }_{i}}        \label{e.gwr}
\end{equation}
Where ${{y}_{i}}$ and ${{x}_{i1}},{{x}_{i2}},...,{{x}_{ip}}$ are observed values of dependent variable and independent variables ${{x}_{1}},{{x}_{2}},...,{{x}_{p}}$ at the location of $\left( {{u}_{i}},{{v}_{i}} \right)$, which is geospatial coordinates of observation point $i~\left( i=1,2,\ldots ,n \right)$, called a model calibration location. In the context of metro networks, ${{y}_{i}}$ and ${{x}_{i1}},{{x}_{i2}},...,{{x}_{ip}}$ are observed metro ridership and potential influencing factors ${{x}_{1}},{{x}_{2}},...,{{x}_{p}}$ at station  $i~\left( i=1,2,\ldots ,n \right)$ with the geospatial coordinates $\left( {{u}_{i}},{{v}_{i}} \right)$.  ${{\varepsilon }_{i}}$ is the normally distributed error term (with the expected value 0 and a constant variance). ${{\beta }_{k}}\left( {{u}_{i}},{{v}_{i}} \right)$$\left( k=1,2,\ldots ,p \right)$  refers to  $p$ unknown coefficients associated with spatial position pending for estimation.  $p$ is the number of independent variables.
The GWR model at location $i$ in matrix term is:
\begin{equation}
y\left( i \right)=X\left( i \right)\beta \left( i \right)+\varepsilon \left( i \right)     \label{e.GWR}
\end{equation}
The vector of estimation regression coefficients at location $i$ is:
\begin{equation}
\hat{\beta }\left( i \right)={{\left[ {{X}^{T}}W\left( i \right)X \right]}^{-1}}{{X}^{T}}W\left( i \right)y   \label{e.gwrvector}
\end{equation}
Where $X={{[{{X}^{T}}(1);{{X}^{T}}(2);...;{{X}^{T}}(n)]}^{T}}$  is the matrix of independent variables, which typically includes a column of 1s for the intercept. $W\left( i \right)=diag\left[ {{w}_{1}}\left( i \right),\ldots ,{{w}_{n}}\left( i \right) \right]$ is the diagonal weights matrix, which contains the geographical weights in its leading diagonal and 0 in its off-diagonal elements. The weights matrix is calculated for each location $i$ and applies weights to observations $j=1,\ldots ,n$. Here, more weight generally will be placed more emphasis on closer observations to the model calibration location $i$. $y$ is the $n\times 1$  vector of dependent variable values; and $\hat{\beta }\left( i \right)={{\left( {{{\hat{\beta }}}_{i0}},{{{\hat{\beta }}}_{i1}},\ldots ,{{{\hat{\beta }}}_{ip}} \right)}^{T}}$ is the vector of $p+1$ local regression coefficients at location $i$ for $p$ independent variables and an intercept term.

The weights matrix $W\left( i \right)$  is calculated from a kernel function $K(j,i)$. There are many kinds of kernel functions. According to the nature of the bandwidth , kernel functions can be divided into two groups: the fixed kernel function and the adaptive kernel function. Gaussian kernel function is a typical example of the fixed kernel, which was expressed as Equation \eqref{e.weight}.
\begin{equation}
{{w}_{j}}\left( i \right)=\exp \left[ -{{\left( \frac{{{d}_{ij}}}{b} \right)}^{2}} \right]    \label{e.weight}
\end{equation}
where  ${d}_{ij}$  is the distance between the calibration location $i$ and location $j$, and $b$ is the kernel bandwidth parameter.The fixed kernel function assumes that the bandwidth $b$ at each location $i$ is a constant across the study area. If the locations $i$ and $j$ satisfy ${d}_{ij}=0$, ${w}_{j}\left( i \right) = 1$, whereas ${w}_{j}\left( i \right)$ decreases
according to a Gaussian curve as ${d}_{ij}$ increases. However, the weights are nonzero for all data points, no matter how far they are from the location $i$ \citep{brunsdon1996geographically}.  If the sample points are distributed as clusters in the study area but not regularly spaced, it is generally appropriate to allow the kernel to adapt this irregularity by increasing its size when the sample points are distributed sparser and decreasing its size when they are denser. A convenient way of implementing this adaptive bandwidth
specification is to choose a kernel which allows the same number of sample points for each estimation.  A commonly used adaptive kernel function is the bisquare  kernel
function, as follows:
\begin{equation}
{{w}_{j}}\left( i \right)=\left\{ \begin{matrix}
{{[1-{{({{d}_{ij}}/b)}^{2}}]}^{2}}, & if & {{d}_{ij}}<b  \\
0, & if & {{d}_{ij}}\ge b  \\
\end{matrix} \right.\
 \label{e.bisquare}
\end{equation}
where ${w}_{j}\left( i \right) = 1$ at the location $i$ (${d}_{ij}=0$) and  ${w}_{j}\left( i \right) = 0$ at the location $i$ when ${d}_{ij}\ge b$ \citep{Guo2008Comparison}. 
 
Both the Gaussian kernel function and the bi-square kernel function are highly dependent on bandwidth $b$. Thus, the kernel bandwidth $b$ should be estimated firstly to fit the GWR model. The most common method in practice is leave-one-out cross-validation (CV) across all the calibration locations proposed by Cleveland(1979) \citep{William1979Robust}. CV is an iterative process that finds the kernel bandwidth with the lowest associated prediction error of all the dependent variables $y\left( i \right)$ . For each calibration location $i$, it removes the data for observation $i$ in the model calibration at location $i$ and predicts  $y\left( i \right)$ from the other data points leaving $i$ out and the kernel weights associated with the current bandwidth. The Leave-One-Out CV (LOO-CV) function is ${{\min }_{b}}\sum\limits_{i=1}^{n}{[{{y}_{i}}}-{{\hat{y}}_{\ne i}}(b){{]}^{2}}$, where ${{\hat{y}}_{\ne i}}\left( b \right)$  is the fitted value ${y}_{i}$ with location $i$ omitted during the fitting process. The optimal bandwidth $b$ minimizes the CV function.

An alternative to CV in kernel-bandwidth estimation is the Akaike information criterion (AIC), proposed by Fotheringham et al (2002)\citep{Fotheringham2002Geographically}. CV and the AIC are also methods applied in model selection, and more details about AIC and model selection can be referred to Burnham and Anderson, 2002 \citep{Burnham2002Model}. There is little published research on discussing if CV and AIC will calculate an uniform solution or whether one method is preferred under a certain condition.In this paper, we adopt CV method. Actually, the selection of kernel function and bandwidth depends on the specific distribution of sample points in the study area, which has a strong impact on the descriptive and predictive power of GWR models \citep{Guo2008Comparison}.
Finally, with the kernel weights calculated at each calibration location $i$ and the determined bandwidth $b$ from the kernel function $K$, the regression coefficients are estimated at each model calibration location, consequently, the values of dependent variable can be estimated as following:
\begin{equation}
\hat{y}\left( i \right)=X\left( i \right)\hat{\beta }\left( i \right)  \label{e.haty}
\end{equation}
\subsection{Distance Matrix considering Network Structure}
\begin{figure}[H]
	\centering
	\centerline{\includegraphics[width=3.5in]{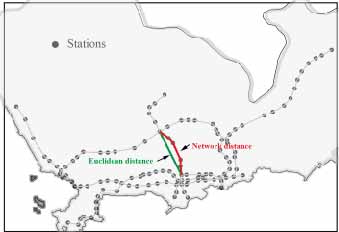}}
	\caption{Sketch map of Euclidean-based distance and Network-based distance. }
	\label{dist}
\end{figure}

In practice, since the GWR model is applied to metro station ridership estimation, it is necessary to take the network connection intermedia into consideration when calculating the distance, however, existing studies usually calculate the distance matrix only considering coordinates, i.e.,just using Euclidean-based distance metric to calibrate the distance no matter how the locations connected. In practice, in metro networks, one station is connected to another station with railway, so the Euclidean-based distance cannot reflect the actual distance between two stations. The study uses network-based distance to measure the spatial connectivity instead of Euclidean-based distance.The two distance measurement methods can be illustarted as Figure \ref{dist}.

The Euclidean-based distance $Ed_{rs}^{{}}$ between stations $r$ and $s$ can be calculated by: 
\begin{equation}
Ed_{rs}^{{}}=R*arccos\left( \cos \left( La{{t}_{s}} \right)*\cos \left( La{{t}_{r}} \right)*\cos \left( Lo{{n}_{s}}-Lo{{n}_{r}} \right)+\sin \left( La{{t}_{r}} \right)*sin(La{{t}_{s}} \right))*\frac{\pi }{180}  \label{e.edist}
\end{equation}
Here, we calculate the distance $Dis{{t}_{i,j}}$ between two close stations  $i$ and $j$ considering the radius of the earth $R$ according to Equation \eqref{e.sdist}:
\begin{equation}
Dis{{t}_{i,j}}=R*arccos\left( \cos \left( La{{t}_{j}} \right)*\cos \left( La{{t}_{i}} \right)*\cos \left( Lo{{n}_{j}}-Lo{{n}_{i}} \right)+\sin \left( La{{t}_{i}} \right)*sin(La{{t}_{j}} \right))*\frac{\pi }{180}  \label{e.sdist}
\end{equation}
Thus, the network-based distance between stations $r$ and $s$  $Nd_{rs}^{{}}$ can be calculated by: 
\begin{equation}
Nd_{rs}^{{}}=\underset{k\in {{K}_{rs}}}{\mathop{\min }}\,\sum\limits_{i=1}^{\mathop{{n}_{k}-1}}{Dis{{t}_{i,i+1}}} \label{e.ndist}
\end{equation}
Where $k$ denotes the $k$-th path between stations $r$ and  $s$,and $k\in {{K}_{rs}}$. ${{K}_{rs}}$ is the set of paths between stations $r$ and  $s$. ${n}_{k}$ is the number of stations the  $k$-th path passes through. So the network-based distance $Nd_{rs}^{{}}$ is exactly the distance of the shortest path along the railway lines, which considers the network connection intermedia. In this way, the $n\times n$ interpoint network-based distance matrix  $\mathbf{D}$ can be obtained.
Therefore, the shortest path length of every two stations considering railway connection will be adopted to calculate the network-based distance matrix. Here, we made use of \textsf{distances} fuction in “igraph” package (see: https://cran.r-project.org/web/packages/igraph/index.html)to calculate the length of all the shortest paths between vertices in the network. As we adjusted the distance measurement of metro networks, the weights calculated from kernel functions will be changed accordingly. 
\subsection{Least Absolute Shrinkage and Selection Operator (LASSO)}
To address the issue of collinearity in the GWR framework,  the Least Absolute Shrinkage and Selection Operator (LASSO) is deemed a proper method because it more directly reduce the variance in the regression coefficient while retaining interpretability of covariate effects\citep{wheeler2005multicollinearity}.In this study, we will also consider the lasso version of GWR framework. As a kind of shrinkage method,  the lasso  not only tends to reduce the variability of the estimates thus can improve the model’s stability, but also can set some of the coefficients to zero, thus allows for variable selection. The lasso makes use of the  $\ell 1$ norm.  $\ell 1$ penalties are convex and the assumed sparsity can lead to significant computational advantages. The lasso is defined as following: 
\begin{equation}
{{\hat{\beta }}^{R}}=arg\underset{\beta }{\mathop{\min }}\,\underset{i=1}{\overset{n}{\mathop \sum }}\,{{({{y}_{i}}-{{\beta }_{0}}-\underset{k=1}{\overset{p}{\mathop \sum }}\,{{x}_{ik}}{{\beta }_{k}})}^{2}}      \label{e.lasso}
\end{equation}
subject to
\begin{equation}
\underset{k=1}{\overset{p}{\mathop \sum }}\,\left| {{\beta }_{k}} \right|\le s,   \label{e.lassoconstrain}
\end{equation}
where $s$ is a parameter that controls the degree of coefficient shrinkage. Tibshirani (1996) proved that the lasso constraint $\underset{k}{\mathop \sum }\,\left| {{\beta }_{k}} \right|\le s$ is equivalent to adding the penalty term  $\lambda ~\underset{k}{\mathop \sum }\,\left| {{\beta }_{k}} \right|$ to the Residual Sum of Squares (RSS). Thus a direct relationship between $s$ and $\lambda \ge 0$ which is a complexity parameter that controls the degree of shrinkage of coefficients. Hence, the lasso coefficients can be also expressed as:
\begin{equation}
{{\hat{\beta }}^{R}}=arg\underset{\beta }{\mathop{\min }}\,\{\underset{i=1}{\overset{n}{\mathop \sum }}\,{{({{y}_{i}}-{{\beta }_{0}}-\underset{k=1}{\overset{p}{\mathop \sum }}\,{{x}_{ik}}{{\beta }_{k}})}^{2}}+\lambda \underset{k=1}{\overset{p}{\mathop \sum }}\,\left| {{\beta }_{k}} \right|\}   \label{e.lassocoefficients}
\end{equation}
The generally used methods for solving the lasso are standard convex optimizer \citep{Gauraha2018Introduction}  and least angle regression (LARS) (Efron et al 2004)\citep{Efron2004Least}. In this study, we adopted LARS to solve the lasso.  The procedure of the LARS algorithm  can be referred to Wheeler, 2009\citep{wheeler2009simultaneous}.
\subsection{Adapted Geographically Weighted Lasso (Ada-GWL) model}
To implement the lasso in GWR, Wheeler (2009) developed a framework called Geographically Weighted Lasso (GWL) and wrote an R package ``gwrr" including \textsf{gwl.est} function to run GWL models \citep{gwrr}. Based on to the framework of Wheeler (2009), we developed our Ada-GWL model in the context of metro ridership estimation by taking network connection intermedia into consideration\citep{wheeler2009simultaneous}. Here, we called \textsf{lars} function in ``lars" package \citep{lars} to implement lasso methods and GWmodel package to implement GWR framework \citep{GWmodel}.

For solving the Ada-GWL model we proposed, the algoritms including the sub-routines and the main body are designed. First of all, the sub-routine to find the best local LARS solution at each location by minimizing LOO-CV Root Mean Square Error (RMSE) using LARS is shown in Algorithm~\ref{algo1}. 

\begin{breakablealgorithm}
	\setstretch{1.1}
	\caption{\textsf{LocalLARS}() algorithm for obtaining the best LARS solution at each location by minizing CV RMSE.}
	\label{algo1}
	\begin{algorithmic}[1]
		\REQUIRE~$b$, the bandwidth; $\mathbf{X}$, a set of observed values of independent variables (factors) ${{x}_{1}},{{x}_{2}},...,{{x}_{p}}$ at the location of $\left( {{u}_{i}},{{v}_{i}} \right)$; $\mathbf{y}$, a set of observed values of dependent variable(ridership); $\mathbf{D}$, the distance matrix; $n$, the number of observation points(stations); $K$, the kernel function.\
		\ENSURE~${{s}_{i}}$, the shrinkage parameter; $\mathbf{z}$, the indicator vector of which variable coefficients are shrunken to zero; $CVerr$, the minimum CV RMSE.\
		\FOR{each station $i,i=1,\ldots ,n$}

		\STATE  Calculate  $\mathbf{W}(i)$ using  $\mathbf{D}$ and  $b $ according to Equation \eqref{e.weight};
		\STATE ${{\mathbf{W}}^{1/2}}\left( i \right)\gets$ $\sqrt{\mathbf{W}(i)}$, and${{\mathbf{W}}^{1/2}}{{\left( i \right)}_{ii}}\gets 0$;
		\STATE  ${{X}_{w}}={{\mathbf{W}}^{1/2}}\left( i \right)\mathbf{X}$ , and ${{y}_{w}}={{\mathbf{W}}^{1/2}}\left( i \right)\mathbf{y}$   ${{\mathbf{W}}^{1/2}}\left( i \right)$ at station $i$;
		\STATE Call function \textsf{lars}$({{X}_{w}},{{y}_{w}})$;

		\RETURN $Parameters({{s}_{i}},\mathbf{z},CVerr)=\underset{parameters}{\mathop{\arg \min }}\,{{\left( {{y}_{i}}-{{{\hat{y}}}_{i}} \right)}^{2}}$
		\ENDFOR
	\end{algorithmic}
\end{breakablealgorithm}

Besides, the sub-routine of bandwidth selection algorithm via binary search is shown in Algorithm~\ref{algo2}. 

\begin{breakablealgorithm}
	\setstretch{1.1}
	\caption{\textsf{BandwidthSelector}() algorithm for selecting the bandwidth $b$ by the binary search for the minimum error.}
	\label{algo2}
	\begin{algorithmic}[1]
		\REQUIRE~$lb$, the lower bound of the initial bandwidth; $ub$, the upper bound of the initial bandwidth; $eps$, loop control value; $\mathbf{X}$, a set of observed values of independent variables (factors) ${{x}_{1}},{{x}_{2}},...,{{x}_{p}}$ at the location of $\left( {{u}_{i}},{{v}_{i}} \right)$; $\mathbf{y}$, a set of observed values of dependent variable(ridership); $\mathbf{D}$, the distance matrix; $K$, the kernel function.\
		\ENSURE~$b$, the bandwidth with the minimum LOO-CV error; $RMSE$, the minimum root mean square error (LOO-CV error).\
		\STATE $diff\gets ub-lb$, $c \gets (lb+ub)/2$,
		\WHILE {$diff>eps$} 
		\STATE $l.c \gets (lb+c)/2$, and $c.u \gets (c+ub)/2$;
		\STATE $l.c.err\gets$ the $CVerr$ returned from \textsf{LocalLARS}$(l.c,\mathbf{X},\mathbf{y},\mathbf{D},n, K)$;
		\STATE $c.u.err \gets$ the $CVerr$ returned from \textsf{LocalLARS}$(c.u,\mathbf{X},\mathbf{y},\mathbf{D},n, K)$; 
		\IF {$l.c.err<c.u.err$}
		\STATE $ub \gets c.u$;
		\ENDIF
		\IF {$l.c.err>c.u.err$}
		\STATE $lb \gets l.c$;
		\ENDIF
		\STATE $c\gets (lb+ub)/2$, $diff\gets |ub-lb|$;
		\ENDWHILE
		\STATE Call the function \textsf{LocalLARS} $(lb,\mathbf{X},\mathbf{y},\mathbf{D},n,K)$, \textsf{LocalLARS} $(ub,\mathbf{X},\mathbf{y},\mathbf{D},n,K)$, and \textsf{LocalLARS} $(c,\mathbf{X},\mathbf{y},\mathbf{D},n, K)$, and compare their $RMSE$ values;
		\RETURN {$b, RMSE$}
	\end{algorithmic}

\end{breakablealgorithm}

With the introduction of sub-routines earlier, the main body for implementing Ada-GWL is shown as Algorithm \ref{algo3}:

\begin{breakablealgorithm}
\setstretch{1.1}
\caption{Pseudo code to estimate the Ada-GWL solutions.}
\label{algo3}	
%\begin{figure}[H]
\begin{algorithmic}[1]
	\REQUIRE$G\left( v,e \right)$, the directed graph generated from a metro network; $\left( {{u}_{i}},{{v}_{i}} \right)$, the geospatial coordinates of observation station $i~\left( i=1,2,\ldots ,n \right)$; $\mathbf{X}$, a set of observed values of independent variables (factors) ${{x}_{1}},{{x}_{2}},...,{{x}_{p}}$ at the location of $\left( {{u}_{i}},{{v}_{i}} \right)$; $\mathbf{y}$, a set of observed values of dependent variable(ridership); $K$, the kernel function.\
	\ENSURE~$b$, the kernel bandwidth parameter; $\hat{\beta }\left( i \right)={{\left( {{{\hat{\beta }}}_{i0}},{{{\hat{\beta }}}_{i1}},\ldots ,{{{\hat{\beta }}}_{ip}} \right)}^{T}}$, the estimated vector of $p+1$ local regression coefficients at station $i$ for $p$ independent variables and an intercept term; ${{\hat{y}}_{i}}$, the fitted value of ridership at each station $i$; $RMSE$, root mean square error; $R-Squared$.\
	\COMMENT{//Phase 1: Estimate the local scaling GWL parameters when the LOO-CV error is minimum. //}
	
	\STATE Calculate the network-based distance $\mathbf{D}$ according to Equation \eqref{e.sdist} and Equation \eqref{e.ndist} ;
	\STATE Set the upper and lower bound of the initial bandwidth $ub\gets ceiling(max(\mathbf{D}))$, and $lb\gets min(\mathbf{D})+0.01\times ub$;
	\STATE $eps \gets RMSE(OLS)\times 0.05$ (or set some other feasible small values) ;
	\STATE ${b,RMSE} \gets \textsf{BandwidthSelector} (lb,ub,eps,\mathbf{X},\mathbf{y},\mathbf{D},K)$; 
	\STATE ${{s}_{i}},\mathbf{z} \gets \textsf{LocalLARS} (b,\mathbf{X},\mathbf{y},\mathbf{D},n,K)$; 
	 \COMMENT{//Phase 2: Estimate the final local scaling GWL solutions using the shrinkage parameter${{s}_{i}}$  and bandwidth  $b$ estimated in Phase 1.//}
	 
	\STATE Using the bandwidth $b$  estimated in Phase 1, call the function \textsf{LocalLARS} $(b,\mathbf{X},\mathbf{y},\mathbf{D},n,K)$;
	\STATE Estimate $\hat{\beta }\left( i \right)$ according to Equation \eqref{e.gwrvector}, and estimate ${{\hat{y}}_{i}}$ according to Equation \eqref{e.haty};
	\RETURN {$b,\hat{\beta }\left( i \right),{{\hat{y}}_{i}},RMSE,R-Squared$}
\end{algorithmic}

\end{breakablealgorithm}
\section{Empirical Study Area and Data}\label{s.scp}
\begin{figure}[H]
	\centering
	\centerline{\includegraphics[width=6.5in]{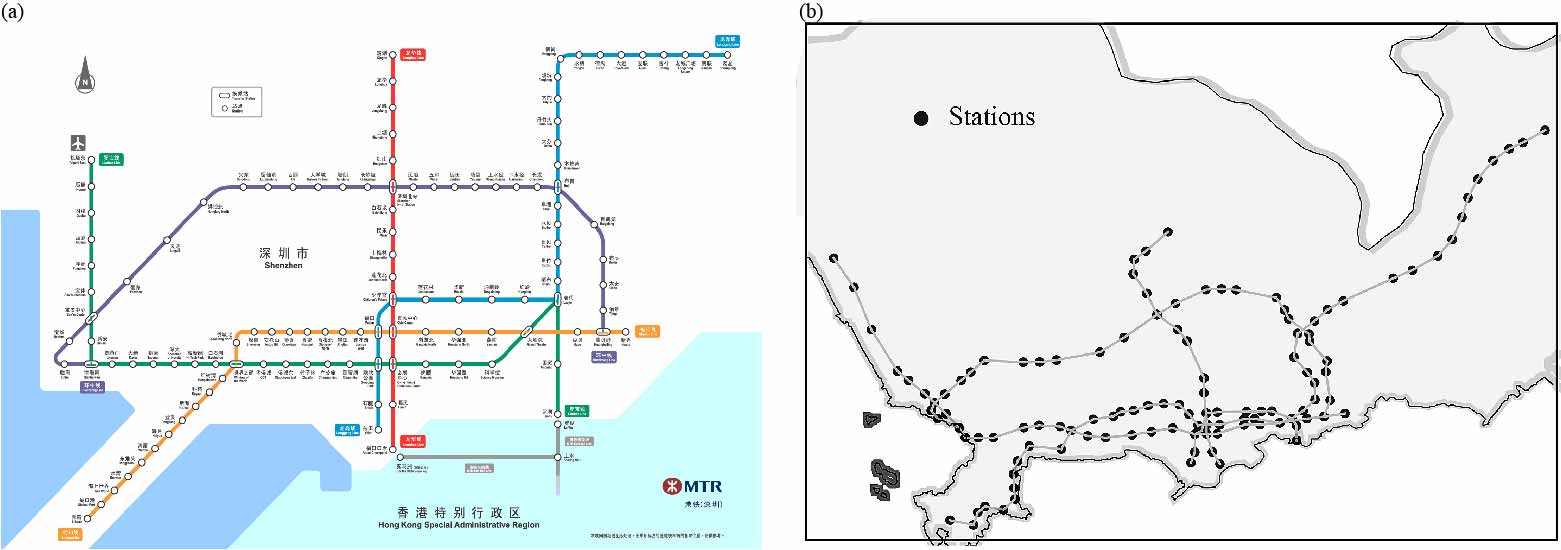}}
	\caption{Shenzhen Metro map of 2013. (a) Schematic map of Shenzhen Metro. (b) Spatial distribution of Shenzhen Metro stations.}
	\label{Fig2}
\end{figure}

In this paper, we concern investigating factors influencing the transit ridership at metro station level in Shenzhen, which is supported by a relatively large Metro transportation network, consisting of 5 lines (Line 1(Luobao Line), Line 2(Shekou Line), Line 3(Longgang Line), Line 4(Longhua Line) and Line 5 (Huanzhong Line)) and 118 stations in the year of 2013.  Figure \ref{Fig2}(a) is the schematic map  of Shenzhen metro in 2013\footnote{Source: \url{http://toursmaps.com/wp-content/uploads/2017/02/shenzhen_metro_map-1.gif}}.  Figure \ref{Fig2}(b) shows the spatial distribution of those stations. The population of Shenzhen at the end of 2013 is about 10,628,900, and the population density is 5,323 persons/km2 \footnote{Source: \url{http://www.sztj.gov.cn/nj2014/indexeh.htm}}. 
\subsection{Data Description}
The Shenzhen metro ridership data at station level were aggregated by using the data collected  through Automatic Fare Collection (AFC) system of Shenzhen Metro Corporation in China. The dataset includes the total information about entry-exit smart card records.  The data used in the research cover a time span of 7 days from October 14th (Mon) to 20th (Sun) in the year of 2013, which includes summed boarding and alighting ridership amounts at two time resolutions with five kinds of time periods: time of the day ((1) rushhour and (2) non-rush hour) and day of the week ((3) average weekday, (4) average weekend and (5) average daily ridership of the whole week). The ridership data at two time resolutions are adopted as dependent variables in the fitting of models. The independent variables represent factors hypothesized to influence station ridership (A detailed description is provided in Table\ref{tab:2}). These variables can be classified into four categories: (1) Social economic variables; (2) Land use variables; (3) Intermodal traffic access variables; and (4) Network structure variables.
Before collecting the data such as land use-related variables and population, a critical first step is to evaluate the walking distance to metro stations, i.e., the size of a PCA, aiming to determine the range of data collection. As the average friendly walking distance is generally assumed to be 500m in large and middle-sized cities according to Kim et al. (2017) \citep{kim2017mapping}, we also define the distance of PCA of each Shenzhen metro station as 500m.  In our work, we use a buffer to create circular PCA by 500m. Based on the buffer with 500m radius determined, population, all of the land use-related data and the number of bus stations are collected subsequently. 

% Table generated by Excel2LaTeX from sheet '工作表1'
\begin{table}[htbp]
	\centering
	\caption{The table of independent variables summary}
	\scalebox{0.70}[0.70]{%
	\begin{tabular}{p{16.25em}p{18.55em}p{11.3em}p{12.35em}}
		\toprule
		\toprule
		\textbf{Categories} & \textbf{Independent variables} & \textbf{Acronym of variables} & \textbf{Source} \\
		\midrule
		\multirow{8}[2]{*}{\textbf{Land Use}} & No. of residential units & \textit{Residence} & Baidu map \\
		\multicolumn{1}{c}{} & No. of restaurants & \textit{Restaurant} & Baidu map \\
		\multicolumn{1}{c}{} & No. retailers/shopping & \textit{Shopping} & Baidu map \\
		\multicolumn{1}{c}{} & No. of schools & \textit{School} & Baidu map \\
		\multicolumn{1}{c}{} & No. of offices & \textit{Offices} & Baidu map \\
		\multicolumn{1}{c}{} & No. of banks & \textit{Bank} & Baidu map \\
		\multicolumn{1}{c}{} & No. of hospitals & \textit{Hospital} & Baidu map \\
		\multicolumn{1}{c}{} & No. of hotels & \textit{Hotel} & Baidu map \\
		\midrule
		\multirow{3}[2]{*}{\textbf{Network Structure}} & Distance to the city center & \textit{Dis\_to\_center} & Calculated \\
		\multicolumn{1}{c}{} & Degree centrality & \textit{Degree} & Calculated \\
		\multicolumn{1}{c}{} & Betweenness centrality & \textit{Betweenness} & Calculated \\
		\midrule
		\multirow{2}[2]{*}{\textbf{Social Economics}} & Population & \textit{Pop} & Worldpop \\
		\multicolumn{1}{c}{} & Days since opened & \textit{Days\_open} & Wikipedia \\
		\midrule
		\textbf{Intermodal Traffic Access} & No. of bus stations & \textit{Bus} & Baidu map \\
		\bottomrule
		\bottomrule
	\end{tabular}%
	\label{tab:2}%
}
\end{table}%

\subsection{Dependent variable}
This paper aims to identify and analyze different factors influencing the ridership at station level. In all available days(Oct. 14th- Oct. 20th), the count of average daily records of weekdays is about 2,513,330, and that of weekends is about 2,480,449, which is less than that of weekdays, indicating the daily trip frequency of weekdays is higher than that of weekends.

\begin{figure}[H]
	\centering
	\centerline{\includegraphics[width=6.5in]{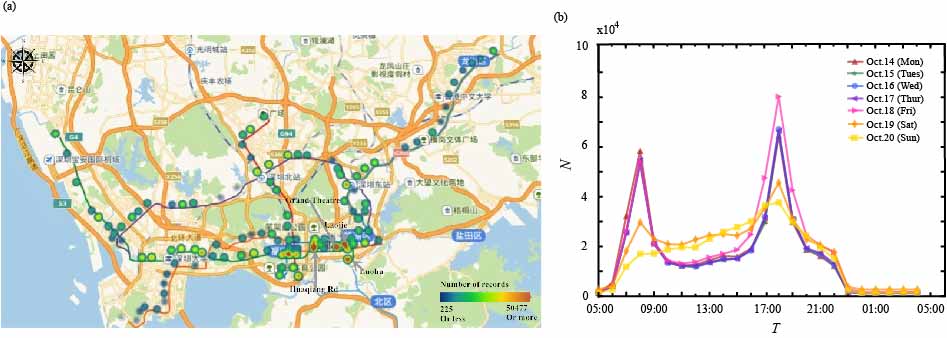}}
	\caption{Spatial and temporal distribution of AFC data records. (a) Spatial distribution of records in one day. (b) Temporal distribution of records per hour in one day.}
	\label{Fig3}
\end{figure}

We conduct preliminary statistical analysis using metro AFC data on October 14. Figure\ref{Fig3}(a) shows the spatial distribution of AFC data records in one day. It presents that the records are most densely distributed at Grand Theatre Station and Laojie Station, closely followed by Huaqiang Road station and Luohu station, and the records of other stations have relatively sparse distribution. 

Figure\ref{Fig3}(b) is about temporal distribution of AFC data records, which shows that the spatial distribution of records has a peak value at both 8:00 and 18:00 on no matter weekdays or weekends, and the records in rush hours on weekends are significantly less than those on weekdays. In addition, it is found that the peak value in the morning rush hour is generally less than that in the evening rush, which may result from people's travel time (such as commuting time) inconsistency in the morning while consistency in the evening. Besides, the records of other periods except for rush hours of weekends are more than those of weekdays. A possible reason could be that people take more non-commuting trips on weekends than on weekdays.

Additionally, the characteristics of temporal distribution of records on weekdays and weekends are quite similar, which suggests that there are similar metro travel patterns with morning and evening peaks on weekdays and weekends in Shenzhen. Moreover, from the perspective of distribution of records in a single day, it is also found that the maximum evening peak value falls on Friday. It indicates that more people tend to take trips in addition to daily commutes on Friday evening because of the last workday.

According to the analysis above, the travel demands and travel patterns not only differentiate in spatial but also differentiate at different time resolutions, such as the level of time of the day and day of the week. 
\subsection{Independent variables}

The independent variables represent factors hypothesized to influence station ridership (Table1). The variables can be classified into four categories: (1) Land use; (2) Social economic variables; (3) Intermodal traffic access variables; and (4) Network structure variables.
\subsubsection{Land use variables}\label{sss.stab_metric}
All of the land use-related data within a PCA were collected from Baidu Map with the assistance of API, and land use variables consist of the stations' nearby residence, entertainment, services, business, education, offices. Specifically, the information covers the numbers of residence, restaurants, schools, offices, hospitals, banks, shopping places, and hotels within 500m PCA. 
\subsubsection{Social economics}
\begin{figure}[H]
	\centering
	\centerline{\includegraphics[width=3.5in]{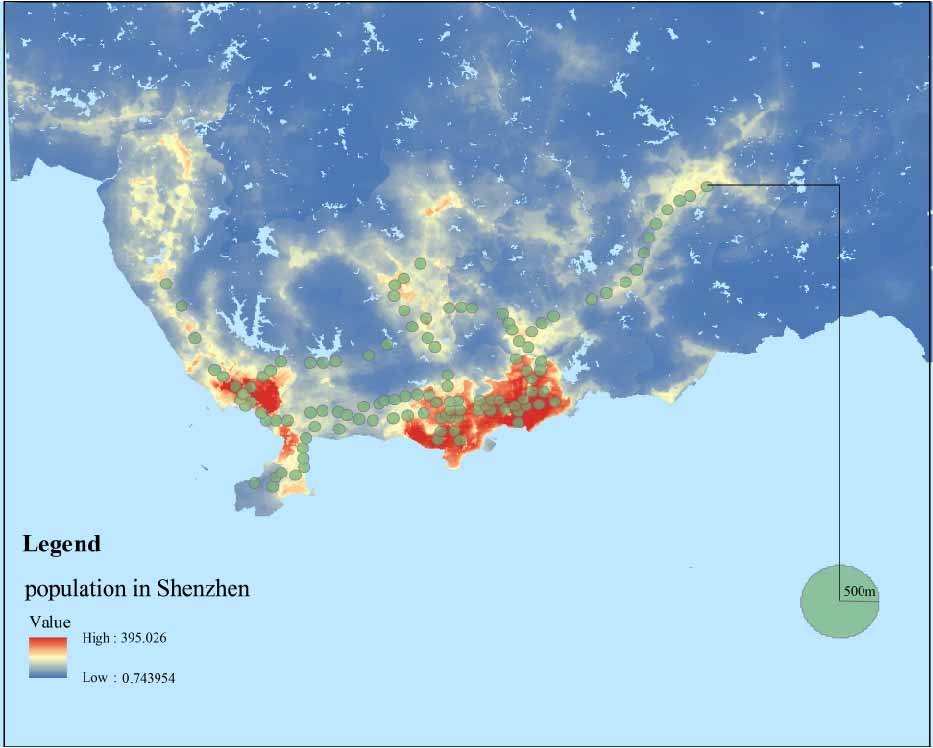}}
	\caption{Population distribution and 500 m buffers of metro stations.}
	\label{Fig4}
\end{figure}

With regard to social economic variables, they consist of the population distribution of Shenzhen in the year of 2013 and operation days since metro stations opened. The information of days since metro lines and stations opened was collected from “Wikipedia” \footnote{Source: \url{https://en.wikipedia.org/wiki/Shenzhen_Metro}} . The population data were collected from the website of Worldpop \footnote{Source: \url{http://www.worldpop.org.uk/data/get_data/}}. The format of the population file is Geotiff. The file provides estimated numbers of people per grid square at $8.33\times {{10}^{-4}}$ degrees spatial resolution (approximately 100m at the equator), which can be projected to ``GCS-WGS-1984" geographic coordinate system. During the data pre-processing, the population within each buffer can be obtained by summing up the value of the grid falls into the metro station buffer. 
Figure \ref{Fig4} shows the population distribution of the whole city of Shenzhen in the year of 2013. Meanwhile, the buffers of metro stations with a radius of 500 m were created by using  ArcGIS 10.2\citep{ArcGIS}, which are also illustrated in Figure \ref{Fig4}. Through the preliminary visualization in Figure \ref{Fig4}, it was noted that population is densely distributed near the metro region. The influence of population density within each station buffer on ridership is pending for analysis in the model.
\subsubsection{Intermodal traffic access variables}
As for intermodal traffic access, here we only considered the feeder bus system, and the number of bus stations within 500m PCA of a metro station was hypothesized to be positively related to station ridership, which was also collected from Baidu Map \footnote{Source: \url{https://map.baidu.com/}}.
\subsubsection{Network structure variables}
In this paper, network structure variables comprise the degree centrality and betweenness centrality of the metro network nodes, and spatial characteristics of each station: distance to city center. In the field of complex networks, as degree is a simple centrality measure that counts how many neighbors a node has, and the betweenness centrality for each node refers to the number of shortest paths that pass through the node (Erciyes, 2015) \citep{Erciyes2014Complex}, thus they are correlated to the information for transfer stations or terminal stations, and the importance of stations in the aspect of their controlling over flows passing between others of metro networks. As for the distance $Dis{{t}_{i}}$  of each station i to the city center, i.e., Shenzhen Municipal People's Government, located in Futian District, we calculate it by the following \eqref{e.dist}  considering the effect of the radius of the earth:
\begin{equation}
Dis{{t}_{i}}=R*arccos\left( \cos \left( La{{t}_{0}} \right)*\cos \left( La{{t}_{i}} \right)*\cos \left( Lo{{n}_{0}}-Lo{{n}_{i}} \right)+\sin \left( La{{t}_{i}} \right)*sin(La{{t}_{0}} \right))*\frac{\pi }{180}              \label{e.dist}
\end{equation}

Where, $R$  is the radius of the earth, $(La{{t}_{0}},~Lo{{n}_{0}})$ and $(La{{t}_{i}},~Lo{{n}_{i}})$ are the latitude and longitude of the city center and station $i$, respectively. The related geographical data were collected from Google map \footnote{Source: \url{https://maps.google.com}}.

\section{Model Implementation and Results Discussion}
In this section, we present the implementation details for metro station ridership modeling using the proposed Ada-GWL. We then use this implementation in our case study for results comparisons (with OLS, GWR, and GWL models), aiming to demonstarte the superiority of Ada-GWL model, and further analyze the results of Ada-GWL combined with the practical meanings.
\subsection{Implementation}
\subsubsection{Data processing}
First of all, for the sake of convenience in representing and understanding, we use alphanumeric code instead of Chinese to denote each station name. Here, we define identifiers for station names according to the following rules: (1) non-transfer stations consist of 3 digits, where the first digit denotes the line number, and the rest 2 digits denote the sequential number of station; (2) transfer stations start with character \textit{t} followed by 3 digits, where the first 2 digits denote the intersection of two lines, and the last digit means the sequential number of intersections between those two lines. For example, ``402" represents the 2nd station of Line 4, and ``\textit{t}131" represents the transfer station that is the first intersection of lines 1 and 3. In this way, all of 118 stations can be encoded by such identifiers containing the line and station information literally.

Shenzhen metro implements AFC system in the whole rail network, which brings chance to record accurately the information of passengers’ boarding and alighting points. The information makes it possible to possess the temporal and spatial distribution law of ridership, getting the real-time ridership data of  each station. This paper is based on the boarding and alighting ridership (smart card) data of 118 stations of Shenzhen metro covering October 14th to October 20th (a whole week), 2013. In the study, we sum up the boarding (entry) and alighting (exit) records at different time resolutions, such as the level of time of the day and day of the week. Therefore, different regression models with different dependent variables, those are rush-hour(evening:17:00-19:00) ridership, non-rush hour (9:00-17:00, 19:00-23:00) ridership, average weekday ridership, average weekend ridership and average daily ridership of the whole week (the operation times of Shenzhen metro is 6:30-23:00), will be built intending to validate the model's effectiveness and applicability to different time resolutions, and further to find if the factors influencing the station-level ridership differentiate at different time resolutions \footnote{Average daily ridership of a whole week is the average of total ridership of seven days of week in operation times (6:30-23:00). Rush hour ridership is calculated by the total ridership of evening rush-hours from 17:00 to 19:00 of a whole week divided by 14 hours (multiply 2 hours by 7 days). Non-rush hour ridership is calculated by the total ridership of remaining time (9:00-17:00, 19:00-23:00) except for morning (7:00-9:00) and evening rush-hours of a whole week divided by 84 hours (multiply 12 hours by 7 days).}.

\begin{figure}[H]
	\centering
	\centerline{\includegraphics[width=6.5in]{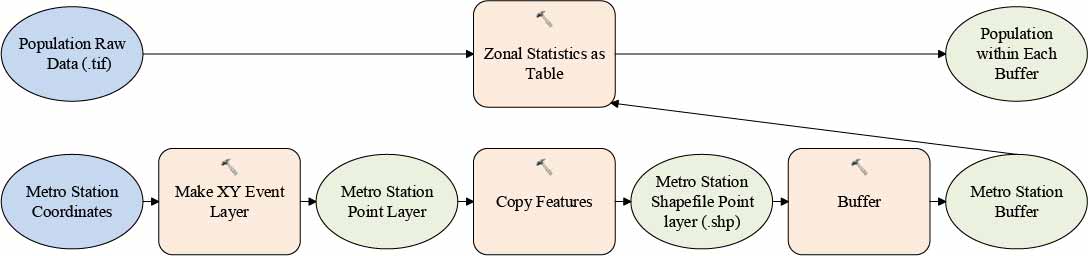}}
	\caption{Procedures of population data processing.}
	\label{poppro}
\end{figure}

For the acquisition of data as independent variables mostly relies on the collection with the assistance of Baidu map API, all of land use variables are the aggregated data collected within 500m PCA, which can be used in the model directly.For the raw population data with Tiff format, they are processed and aggregated within 500m buffers with ArcGIS 10.2 . 
Figure \ref{poppro} shows the overall procedures of population data processing with ArcGIS 10.2. The processing procedure contains 3 main steps described as follows. 
\begin{enumerate}[itemindent=1.5em,label=\textbf{Step} \arabic*:]

\item Transform the coordinate information of metro station into points with shapefile format. Create a new point feature layer based on x and y coordinates defined in a source table and geographic coordinate system by using “Make XY Event Layer” tool.
\item Generate 500m buffers of metro stations. Copy features from the input layer to a feature class (shapefile format) by using “Copy features” tool as the layer above is temporary. Then Create buffer polygons around input features to 500 meters by using “buffer” tool.
\item Zonal statistical analysis. Aggregate the population within each buffer. Since the value of each grid represents the number of people per grid square at $8.33\times {{10}^{4}}$ degrees spatial resolution (approximately 100m at the equator), so the population within each buffer can be obtained by summing up the value of the grid falls into the buffer by using “Zonal Statistics as Table” tool.
	 
\end{enumerate}

\subsubsection{Spatial autocorrelation test to variables}
\begin{figure}[H]
	\centering
	\centerline{\includegraphics[width=6.5in]{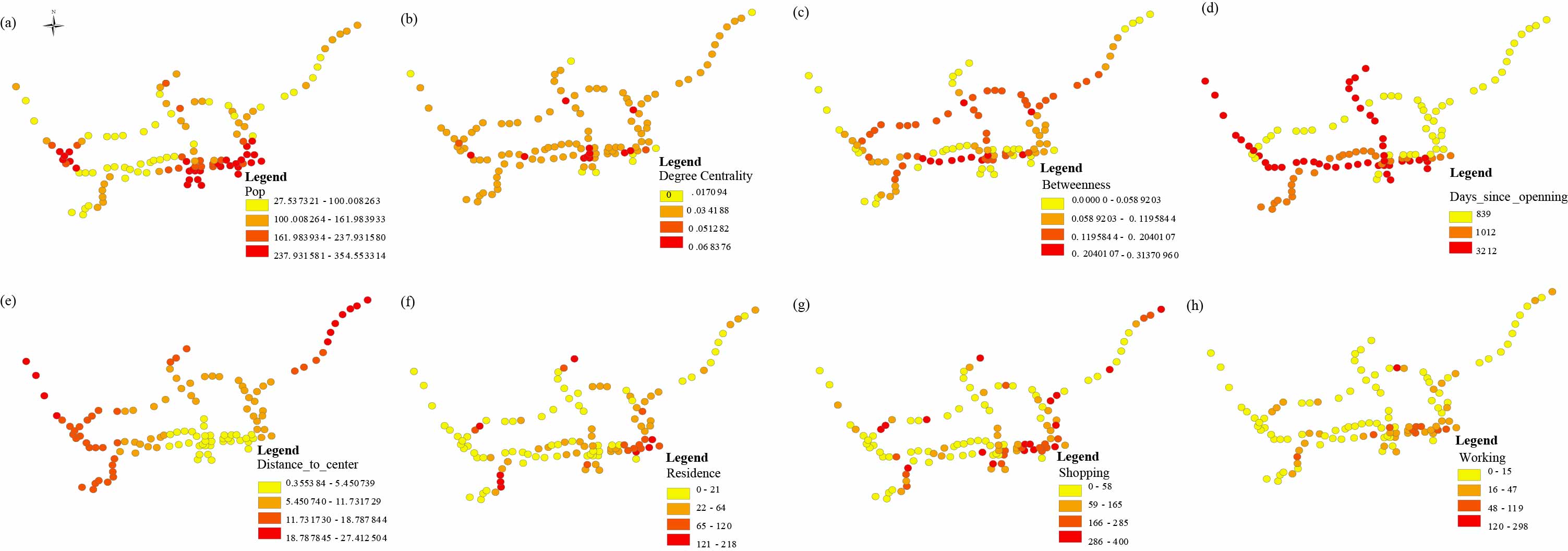}}
	\caption{ spatial distribution of independent variable candidates.}
	\label{Fig5}
\end{figure}

\begin{figure}[H]
	\centering
	\centerline{\includegraphics[width=6.5in]{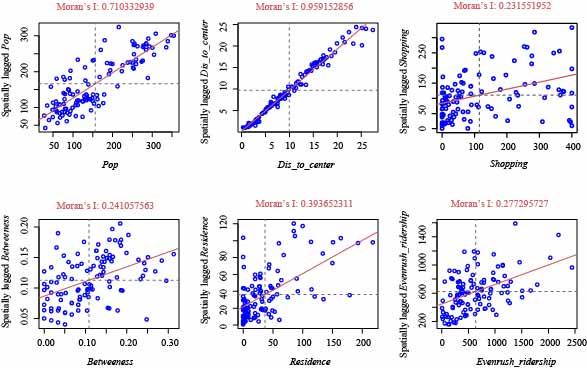}}
	\caption{Moran scatterplot of some variables.}
	\label{Fig6}
\end{figure}

Before building geographically weighted models, analysis was performed to test whether the candidate variables were spatially autocorrelated. First of all, the spatial distribution of variable candidates is analyzed. The results  (Figure \ref{Fig5})  indicate that these independent variables do have certain spatial correlation and aggregation to some extent. For example, population is densely distributed in the southeast regions, covering Luohu district and Futian district. Areas with large betweenness centrality generally locate on the closed railway loop of the metro network. Besides, the openning time of stations of Shenzhen metro is associated with the whole line openning time, and the distance to city center is exactly calculated according to the geographical location information, so these two variables have significant spatial correlation.
With the overview of spatial distribution of independent variables, the test of spatial autocorrelation can further detect how strong spatial correlation of variables are, this will provide a theoretical basis for the feasibility of applying a GWR model. Moran's I is a measure of spatial autocorrelation developed by Patrick Alfred Pierce Moran (1950) \citep{Moran1950NOTES} . All of the dependent variables and candidate predictors had estimated Moran's I values higher than the expected E(I) (Table \ref{table2}), thus, a positive spatial autocorrelation existed . Moran scatter plot can reflect the spatial autocorrelation intuitively. The scatter plot has four quadrants. If the observed value falls to the first and third quadrants, it indicates that there is a strong positive spatial correlation. If it falls to the second and fourth quadrants, it indicates there is a strong negative spatial correlation. Figure \ref{Fig6} shows several variables' Moran scatter plot. 

% Table generated by Excel2LaTeX from sheet 'Sheet1'
\begin{table}[htbp]
	\centering
	\caption{Moran's I tests for spatial autocorrelation on the dependent variables and some of the candidate predictors.}
	\begin{tabular}{cp{10.51em}ccc}
		\toprule
		\toprule
		& \textbf{Variable} & \multicolumn{1}{p{5.255em}}{\textbf{Moran's I}} & \multicolumn{1}{p{6.255em}}{\textbf{Expected I}} & \multicolumn{1}{p{4.155em}}{\textbf{p-value}} \\
		\midrule
		\multicolumn{1}{c}{\multirow{5}[1]{*}{Dependent variables}} & Weekly\_ridership & 0.24116622 & -0.00854701 & 1.21E-06 \\
		& Weekday\_ridership & 0.26389365 & -0.00854701 & 1.44E-07 \\
		& Weekend\_ridership & 0.16612276 & -0.00854701 & 0.0004006 \\
		& Evenrush\_ridership & 0.27729573 & -0.00854701 & 3.90E-08 \\
		& Nonrush\_ridership & 0.24779725 & -0.00854701 & 4.64E-07 \\
		\multicolumn{1}{c}{\multirow{14}[1]{*}{Candidate predictors}} & Residence & 0.39365231 & -0.00854701 & 1.35E-14 \\
		& Restaurant & 0.34171521 & -0.00854701 & 1.69E-11 \\
		& Shopping & 0.23155195 & -0.00854701 & 3.92E-06 \\
		& School & 0.28917679 & -0.00854701 & 4.16E-11 \\
		& Offices & 0.17494683 & -0.00854701 & 1.37E-05 \\
		& Bank & 0.38831017 & -0.00854701 & 1.98E-14 \\
		& Bus & 0.31229433 & -0.00854701 & 7.90E-10 \\
		& Hospital & 0.21127159 & -0.00854701 & 1.26E-05 \\
		& Hotel & 0.46648095 & -0.00854701 & \multicolumn{1}{p{4.155em}}{$<$2.2e-16} \\
		& Dis\_to\_center & 0.95915286 & -0.00854701 & \multicolumn{1}{p{4.155em}}{$<$2.2e-16} \\
		& Degree & 0.12583388 & -0.00854701 & 0.005471 \\
		& Betweenness & 0.24105756 & -0.00854701 & 1.67E-06 \\
		& Pop & 0.71033294 & -0.00854701 & \multicolumn{1}{p{4.155em}}{$<$2.2e-16} \\
		& Days\_open & 0.38318677 & -0.00854701 & 2.11E-13 \\
		\bottomrule
		\bottomrule
	\end{tabular}%
	\label{table2}%
\end{table}%

According to Moran scatter plots (see Figure \ref{Fig6}), it can be seen that Moran's I values of all variables above are not close to 0, which indicates these variables are  not randomly distributed in space, and mostly falls to the first and the third quadrants. It is shown that each variable is positively spatial correlated more or less, especially eight of independent variables: population, distance to city center, and days since opened, and the number of residence, restaurants, banks, bus stations, and hotels have strong spatial correlation as Moran's I is greater than 0.3 (Cressie, 1993) \citep{Cressie1993Statistics}. The result also lays the foundation for the feasibility of follow-up study.
\subsubsection{Local collinearity diagnostics}
\begin{figure}[H]
	\centering
	\centerline{\includegraphics[width=6.5in]{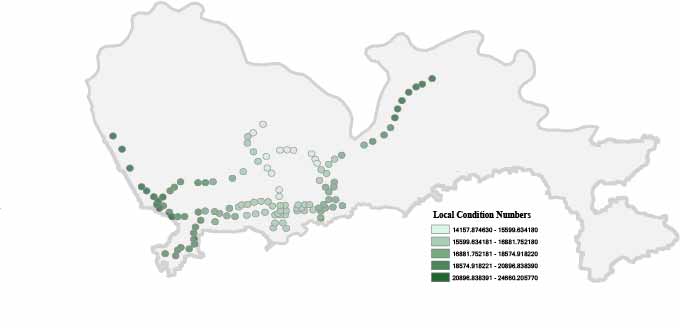}}
	\caption{Spatial distribution of local condition numbers of the GWR model for estimating the average daily ridership of a whole week.}
	\label{Figlocal}
\end{figure}

Since strong spatial correlation was found for the variables in the research, it is reasonable to build GWR models to analyze influencing factors on station ridership of Shenzhen Metro. Here, condition number that evaluated collinearity were 32.50 ($<100$), indicating the low degree of multicollinearity among the candidate independent variables. However, through a series of local collinearity diagnostics for the local regression coefficients of GWR model for ridership estimation at different time resolutions, we found there was severe collinearity in local regression coefficients according to local condition numbers of all stations greater than 1000 (see Figure \ref{Figlocal}) . Multicollinearity is of great concern when condition numbers are greater than 1000, and results become unstable in the presence of strong local collinearity. It just showed that the local regression coefficients were potentially collinear even if the underlying exogenous variables in the data generating process were uncorrelated \citep{wheeler2005multicollinearity}. Due to the collinearity issue in the GWR model for our case, it is proper to implement Ada-GWL model to the data.
\subsubsection{Model implementation}
A series of models icluding OLS, GWR, GWL, and the proposed Ada-GWL were built in  R software\citep{Rsoftware} by using the ``GWmodel", ``gwrr", ``fields", ``lars" and ``igraph packages \citep{GWmodel,gwrr,fields,lars,igraph}.The algorithm for solving the Ada-GWL model is coded in R programming language and tested on Intel(R) Core(TM) i5-6500 CPU, 3.20 GHz processor. Windows 7 operating system is used for testing. For GWR, GWL, and Ada-GWL models,  the selection of kernel function and bandwidth is necessary, which depends on the specific distribution of sample points in the study area, which has a strong impact on the descriptive and predictive power of GWR models \citep{Guo2008Comparison}, it is necessary to consider the distribution of sample points of Shenzhen metro network while selecting the kernel function. Since the stations of Shenzhen metro are relatively regularly spaced, so we adopted Gaussian kernel function to calculate the weight matrix and determined the bandwidth $b$ based on CV.
\subsection{Results analysis}
\subsubsection{Comparative analysis}

In the following part, OLS model, GWR model (an spatial extension of OLS), Adapted  GWR (Ada-GWR) model for metro network (i.e., GWR model calibrated with network-based distance metric), GWL model (GWR with LASSO enabling simultaneous coefficient penalization and model selection), Ada-GWL model for metro network (GWL model calibrated with network-based distance metric) are implemented to the Shenzhen metro dataset described earlier, the results of models for estimating ridership at different time periods are shown in Figure \ref{Figcompari}:

\begin{figure}[H]
	\centering
	\centerline{\includegraphics[width=6.5in]{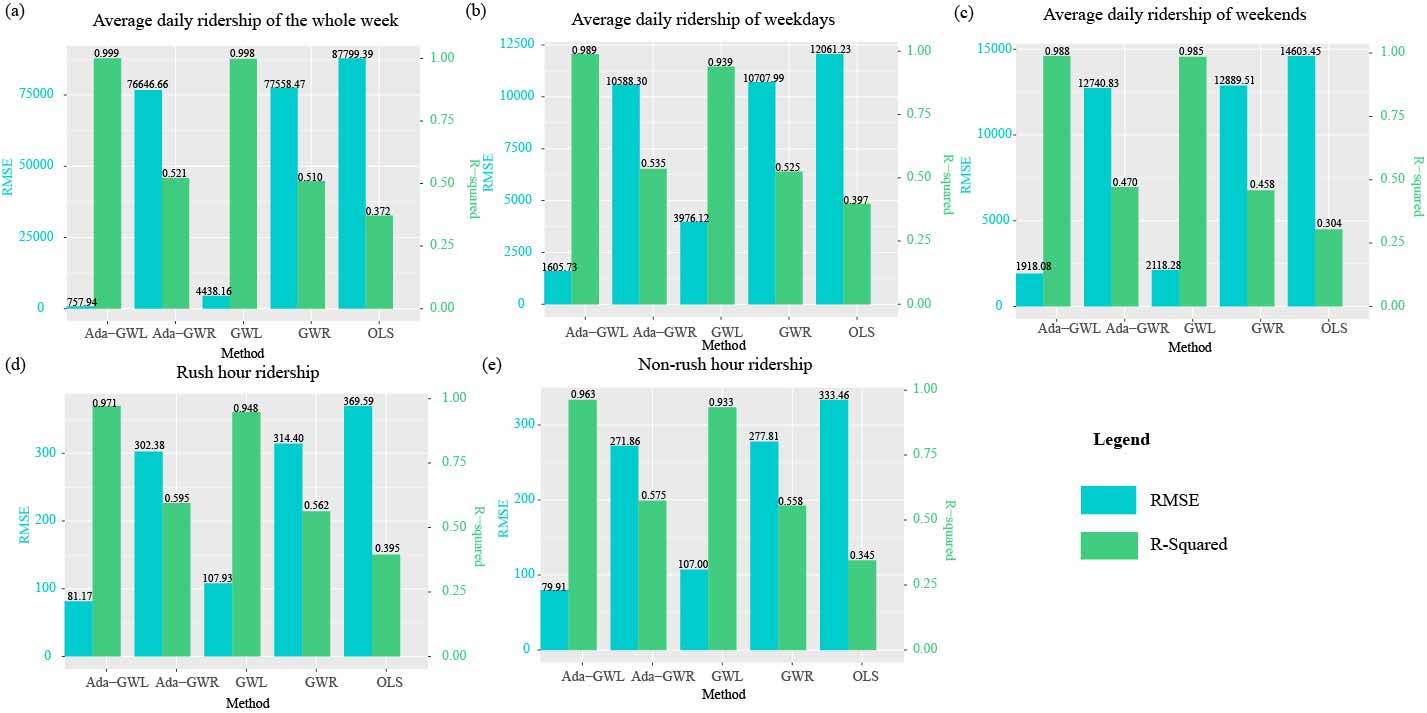}}
	\caption{Comparison of regression performance of models for ridership estimation at different time periods. (a) Average daily ridership of the whole week. (b) Average daily ridership of weekdays. (c) Average daily ridership of weekends. (d) Average hourly ridership of evening rush hours. (e) Average hourly ridership of non-rush hours.)}
	\label{Figcompari}
\end{figure}
% Table generated by Excel2LaTeX from sheet 'Sheet1'

The accuracy of the estimated responses is measured by calculating RMSE. The RMSE is the square root of the mean of the squared deviations of the estimates from the true values, and should be small for accurate estimators. R-squared is a statistical measure that represents the proportion of the variance for a dependent variable that's explained by independent variables. The results of fitting five models to the data provide the error values shown above. In general, the performance ranks of five methods are the same for the five models (different time periods), which are ``Ada-GWL$>$GWL$>$Ada-GWR$>$GWR$>$OLS". First of all, we can see the superiority of four local models (GWR, Ada-GWR, GWL, and Ada-GWL) over global model (OLS) in terms of estimation error of dependent variable and goodness of fit. Secondly, it is noted that the adapted models including both GWR and GWL performed better than original versions of GWR and GWL models, which proved the fitness of the adapted models to the metro network. Thirdly, Ada-GWL for metro network do substantially better than other three models at estimating the dependent variable. Therefore, we can conclude that Ada-GWL model for the metro network we proposed can estimate Shenzhen metro ridership at any time period more accurately.
\subsubsection{Analysis of the local regression of Ada-GWL model}
1)  To see which station a certain variable has the most impact on

\begin{figure}[H]
	\centering
	\centerline{\includegraphics[width=5.5in]{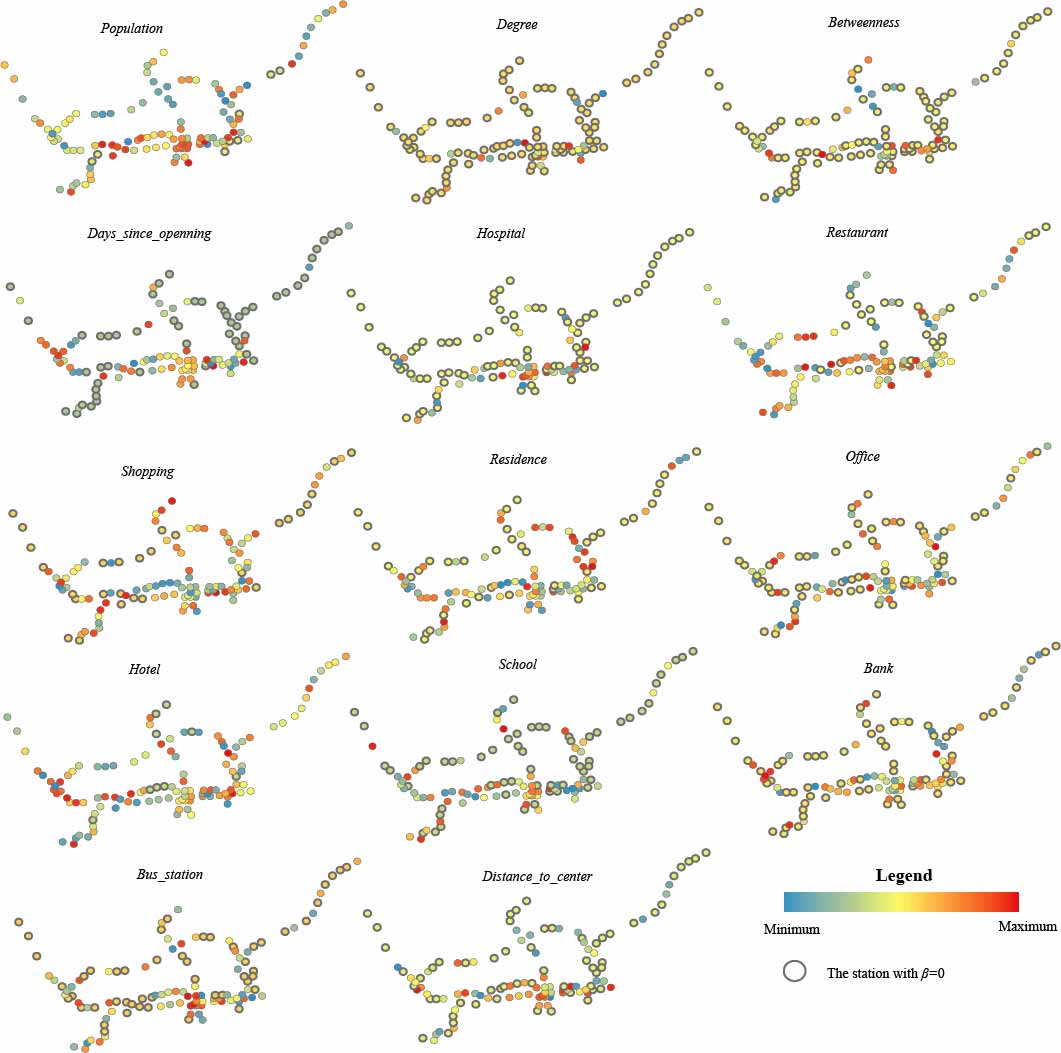}}
	\caption{Spatial distribution of local coefficients (elasticities) of Ada-GWL model for average daily ridership of a whole week.}
	\label{Fig7}
\end{figure}

Focusing on the result of Ada-GWL model, the local regression coefficients' distribution for each variable of all stations can be plotted as the bubble plots. Take the modelfor estimating average daily ridership of a whole weel as an example. The spatial distribution of local coefficients is shown in Figure \ref{Fig7}.
Through understanding the spatial distribution of local coefficients (elasticities), it is possible to know how relations between the variables vary across space (estimated coefficients) and variables selection of all stations. In above figures, the bubble with the bold outline demonstrates the coefficient for the variable in the station equals to 0, and other bubbles’ colours identify the range of coefficients, the bubble with lighter colours mean the coefficient is larger, and vice versa. Take the population factor as an example, stations with large positive coefficients are mainly distributed in the centre, meaning that more trips per capita were expected in the central south area of the metro network, where commerce, administration and education are concentrated. Besides, we can note that for the factors of degree and hospital, there are a large number of stations with zero coefficient, so these factors are not so important factors for influencing most of Shenzhen metro stations' ridership.

\begin{figure}[H]
	\centering
	\centerline{\includegraphics[width=5.5in]{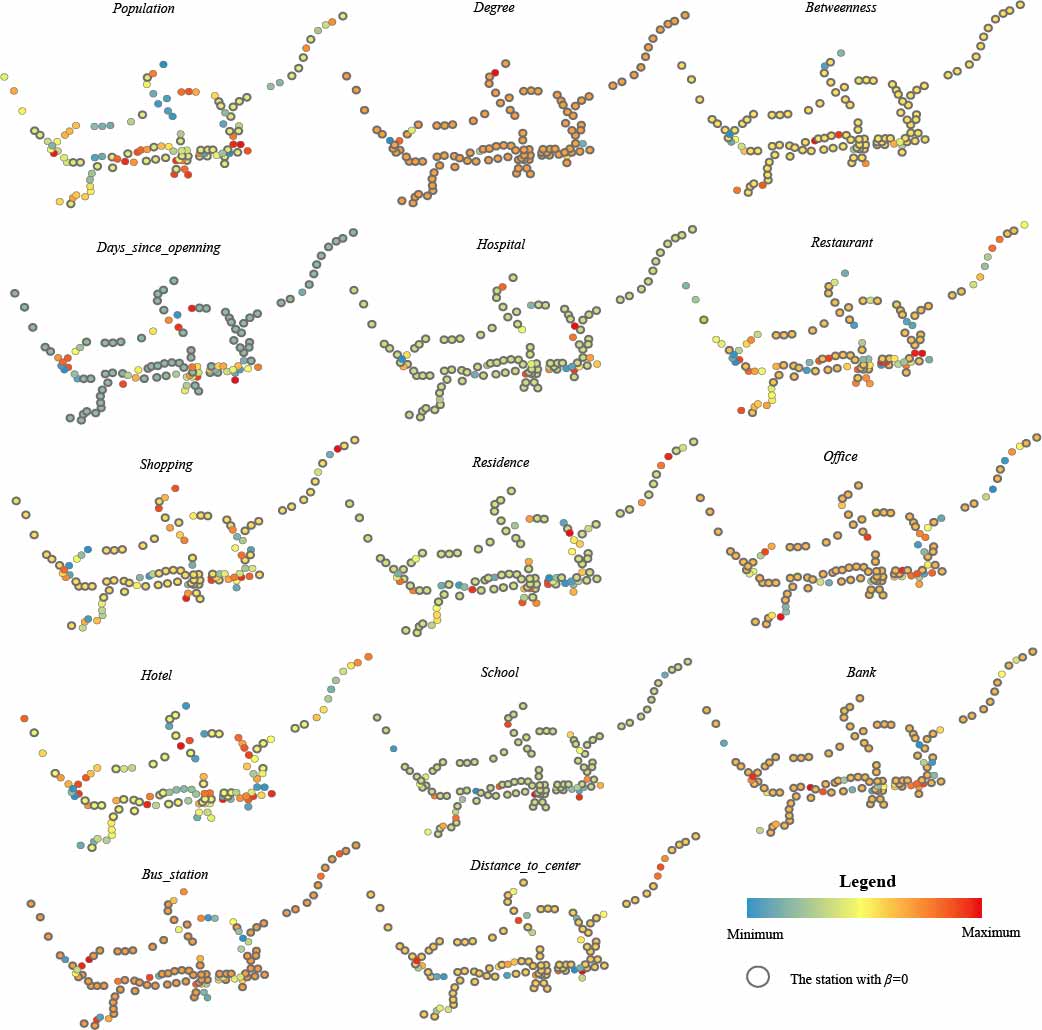}}
	\caption{Spatial distribution of local coefficients (elasticities) of Ada-GWL model for non-rush hour ridership.}
	\label{Fig10}
\end{figure}

Besides, in order to verify the applicability of Ada-GWL model to different time resolutions. We apply the models to estimate the ridership of other time resolutions. With regards to the models for non-rush hour ridership, the spatial distribution of local coefficients is illustrated in Figure \ref{Fig10}.

According to Figure \ref{Fig10}, we can see that more stations with zero coefficient than those of average daily ridership, and the coefficient matrix is very sparse, indicating that for a single station, main factors influencing the hourly ridership at non-rush hour are not manifold.

2) To see which variable has the biggest impact on a certain station

The heat map is applied to demonstrate the coefficient distribution of all variables and all stations (scaled by column (station) at first). For example, for station ``313" (i.e., Tianbei Station), population is the factor that has the biggest positive impact on its ridership compared with other factors. For station ``t251" (i.e., Huangbeiling Station), the number of shopping malls in its PCA has the biggest positive impact on its ridership compared with other factors.

\begin{figure}[H]
	\centering
	\centerline{\includegraphics[width=5.5in]{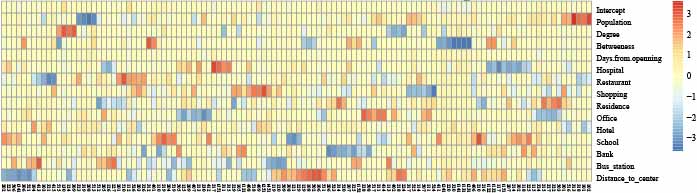}}
	\caption{Heatmap of coefficients of each station for average daily ridership of a whole week.}
	\label{Fig8}
\end{figure}

\begin{figure}[H]
	\centering
	\centerline{\includegraphics[width=5.5in]{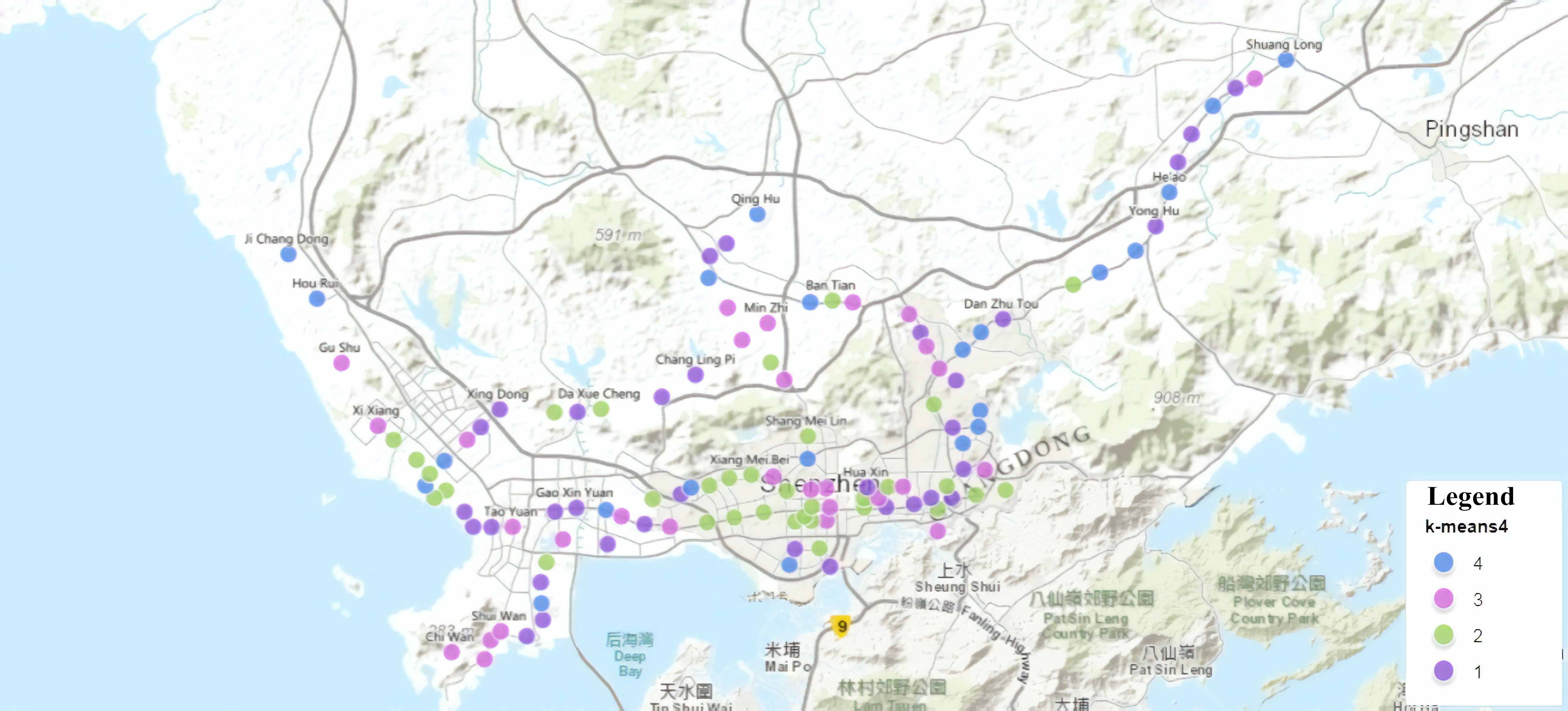}}
	\caption{Clusters by coefficients of stations for average daily ridership of a whole week with the method of k-means.}
	\label{Fig9}
\end{figure}

Furthermore, in order to present the impacts of different factors on different stations more intuitively, based on the coefficients of (scaled by column (station)) of the model for average daily ridership of a whole week, we conduct clustering (Figure\ref{Fig9}) to further illustrate the association among stations, so that we can see which groups of stations that all variables have similar effects on, that is, for a group of stations, some variables have the same great influence on ridership (coefficients are large) and some variables have the same slight influence on ridership (coefficients are small). K-means clustering is one of the simplest and popular unsupervised machine learning algorithms. The goal of K-means clustering is to find groups in the data, with the number of groups represented by the variable $K$. K-means algorithm works iteratively to assign each data point to one of $K$ groups based on the features that are provided. Data points are clustered based on feature similarity. Here, we choose the k-means clustering method in this study. 	We determine the number of groups $K$ as 4  through minimizing the difference of Within-Cluster Sum of Squares (WCSS).

According to Figure \ref{Fig9}, the four clusters are coloured by four colours. Referring to the distribution of functional zones in Shenzhen \footnote{Source: \url{http://www.szgeoinfo.com:5001/msmap/flex/landuse.html}} , we can see that the stations of cluster 1(purple) are mainly located in employment and educational based land. Stations in cluster 2 (green) are mainly located in commercial land. Stations in cluster 3 (pink) mainly distribute in traffic hubs land, and cluster 4(blue) are mainly located in residential based areas. To analyze the rationality of the categories as shown in the cluster analysis and the functional characteristics of the catergories, the results are summarized in Table \ref{tab:4}.

% Table generated by Excel2LaTeX from sheet 'Sheet1'
\begin{table}[H]
	\centering
	
	\caption{Clusters and their functional characteristics}
	\scalebox{0.8}{
	\begin{tabular}{cp{8.5em}cp{21.11em}p{10.055em}}
		\toprule
		\toprule
		\multicolumn{1}{p{2.945em}}{Cluster} & Functional zone & \multicolumn{1}{p{5.055em}}{Cluster size (Number of stations)} & Functional characteristics & Representative stations \\
		\midrule
		1 & Employment and educational based & 34 & The coefficients of offices and schools land-use are relatively larger than those of other variables, so offices and schools can attract more commuting ridership in this zone. The change of offices and schools land-use will have great impact on the ridership of stations in this cluster. & Shenzhen University; Xili; High-tech Park; Daxin; Qianhaiwan; \\
		2 & Commercial land & 33 & The coefficients of shopping land use are relatively larger than other variables, indicating shopping land use has great positive impact on these stations. Surrounding the stations in this cluster, commercial, recreational and other land uses are signally larger than other categories land uses, which attracts a large amount of non- commuting ridership. & Guomao; Huaqiang Rd; Huaqiang North; Shopping Park; Houhai; Convention\&Exhibition Center;  \\
		\multirow{2}[0]{*}{3} & \multirow{2}[0]{*}{Traffic hubs land} & \multirow{2}[0]{*}{29} & The coefficients of degree and bus stations are relatively larger than those of other variables, meaning that the transfer stations and bus stations have great impact on ridership. The stations of this kind are mainly located in urban external & \multicolumn{1}{c}{} \\
		& \multicolumn{1}{c}{} &   & transportation hub (including railway station, passenger transport station, cross-border checkpoint, and Special Economic Zone Port), the station ridership is mainly derived from the passenger flow of traffic hub. & Luohu; Shenzhen North Station; Shekou Port; Chiwan; Buji; Xixiang \\
		4 & Residential based & 22 & The coefficients of residential land use are relatively larger than those other variables, indicating the station ridership is mainly affected by the commuting passenger flow from residential areas. stations of this kind are also mainly located in residential based areas (even though the number of residential units surrounding these stations might be not the largest among other kinds of land use).  & Baishizhou; Wuhe; Bu Xin; Jixiang;  \\
		\bottomrule
		\bottomrule
	\end{tabular}}%
	\label{tab:4}%
	
\end{table}%

By this token, Ada-GWL model can offer clear categorizing of functional zones of Shenzhen metro stations belong to, which makes findings more explainable and instructive to metro transportation and urban planning.  

\begin{figure}[H]
	\centering
	\centerline{\includegraphics[width=5.5in]{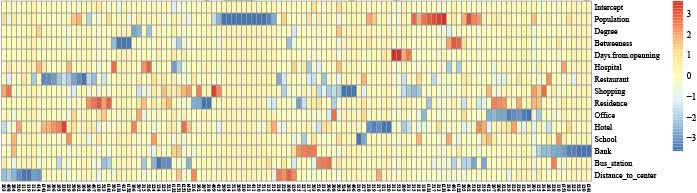}}
	\caption{Heatmap of coefficients of each station for normal time hourly ridership.}
	\label{Fig11}
\end{figure}

\begin{figure}[H]
	\centering
	\centerline{\includegraphics[width=5.5in]{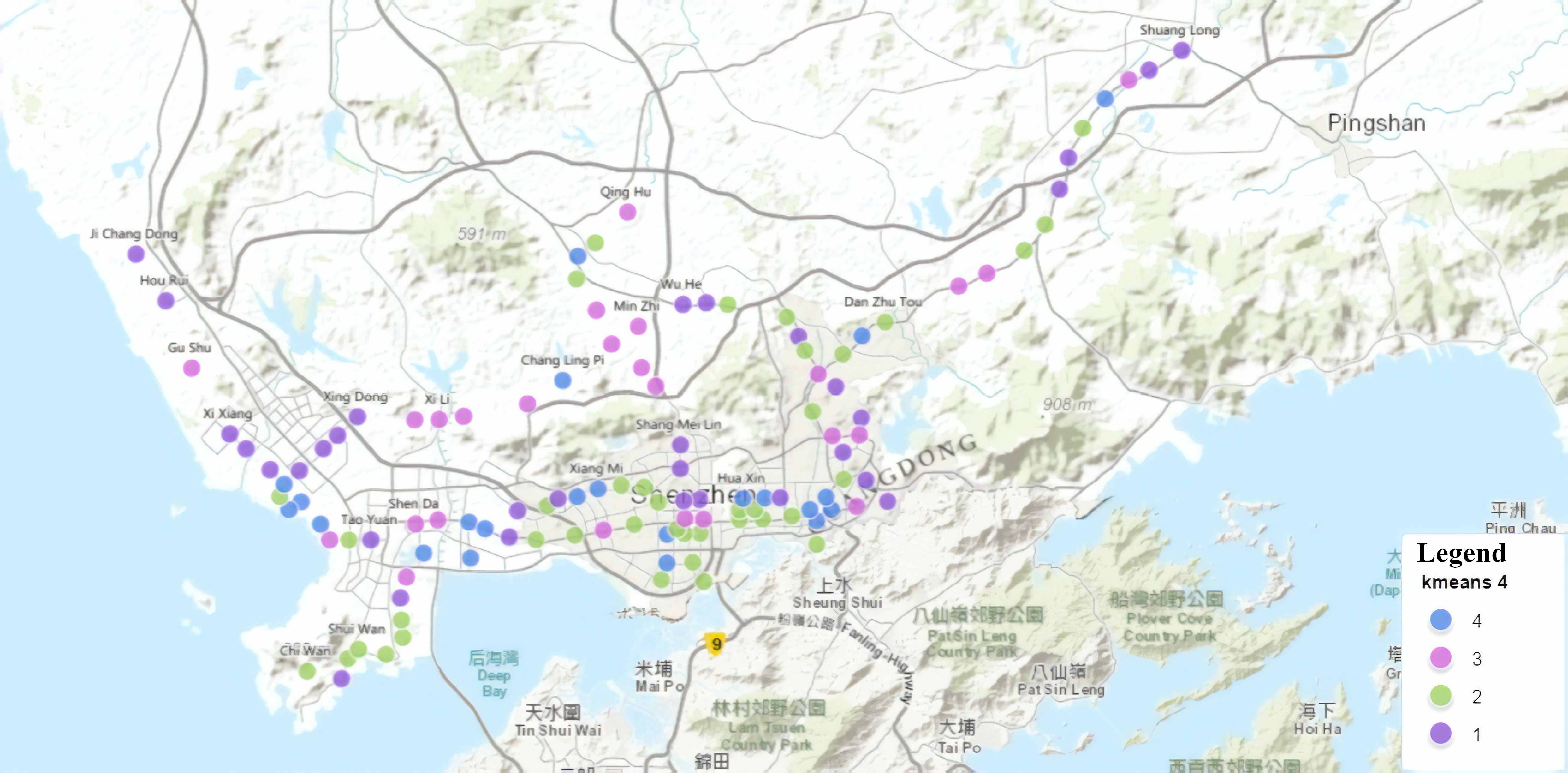}}
	\caption{Clusters by coefficients of stations for non-rush hour ridership with the method of k-means.}
	\label{Fig12}
\end{figure}

Here, we also use heatmap to demonstrate the coefficient distribution of all variables and all stations (scaled by column (station) at first) for non-rush hour ridership (see Figure \ref{Fig11}), also, based on the coefficients of stations of the model for non-rush hour ridership, we did clustering to illustrate the association among stations (see Figure \ref{Fig12}).


According to Figure \ref{Fig12}, stations belong to cluster 1 (green) are mainly located in residential based areas. Stations in cluster 2 (green) are mainly located in commercial land. Stations in cluster 3 (pink) mainly distribute in traffic hubs land, and cluster 4(blue) are mainly located in employment and educational based land. The result is slightly different from that for average daily ridership, indicating different time resolutions will make stations present different characteristics.
\section{Conclusion and Impications for Metro Planning and Periphery Development }\label{s.concl}
In summary, this paper proposed an Ada-GWL framework for modelling the ridership of metro systems, which considered the network connection intermedia when calculating the distance matrix instead of Euclidean-based distance, and also simultaneously performed regression-coefficient shrinkage and local model selection. We demonstrated the superiority of our framework through a real world case study of Shenzhen Metro Systems, and meanwhile, we analyzed the influencing factors of Shenzhen metro station-level ridership from a local perspective. Different from previous works, the proposed direct demand model based on Ada-GWL framework engaged local model selection and network-based distance to estimate the station-level ridership accurately.

The results of the case study showed that Ada-GWL model performed the best in terms of goodness-of-fit compared with OLS, the ordinary GWR model, Ada-GWR calibrated with network-based distance, and ordinary GWL model. The implementation of the proposed Ada-GWL model can provide a theoretical basis for interpreting the influencing factors of station-level ridership from a local perspective. First, through understanding the spatial distribution of local coefficients (elasticities), it was possible to know how relations between the variables vary across space (estimated coefficients) and variables selection of all stations, which was helpful to see  which station an influencing factor has the most impact on. Second, the coefficient distribution of all variables and all stations made it possible to see which variable has the biggest impact on a given station. Third, through clustering analysis of the stations according to the regression coefficients, clusters' functional characteristics were found to be in compliance with the facts of the functional land use policy of Shenzhen. 

In general, the adapted version of GWL introduced in this paper not only extended the traditional GWR to deal with collinearity issues among regression coefficients and estimation of station-level ridership but also inspired  metro planning and periphery development from a local perspective. First, Transit Oriented Development (TOD) planning is suggested to adjust measures to local conditions. Here, local conditions mean more than the geographical factors, and they also incorporate different local effects of influencing factors on metro ridership. Second, different functional zones need to adopt different strategies of diverting passenger flows. For example, in commercial developed areas, enhancing security when diverting passenger flows is vital, while improving efficiency is more important when diverting passenger flows in employment based areas. Third, for stations with unsaturated passenger flows, the attraction to passenger flows can be boosted in combination with the functionality of the region. These findings can also inspire the metro planning and periphery development of other cities.

\appendix 
\section{Appendix} 
\setcounter{figure}{0}
\renewcommand{\thefigure}{A\arabic{figure}}
\subsection{Summary of literature review on direct demand models for ridership estimation}\label{a.litertab}

The related studies on direct demand models for ridership estimation are summarized as following. 
\begin{figure}[H]
	\centering
	\centerline{\includegraphics[width=6.5in]{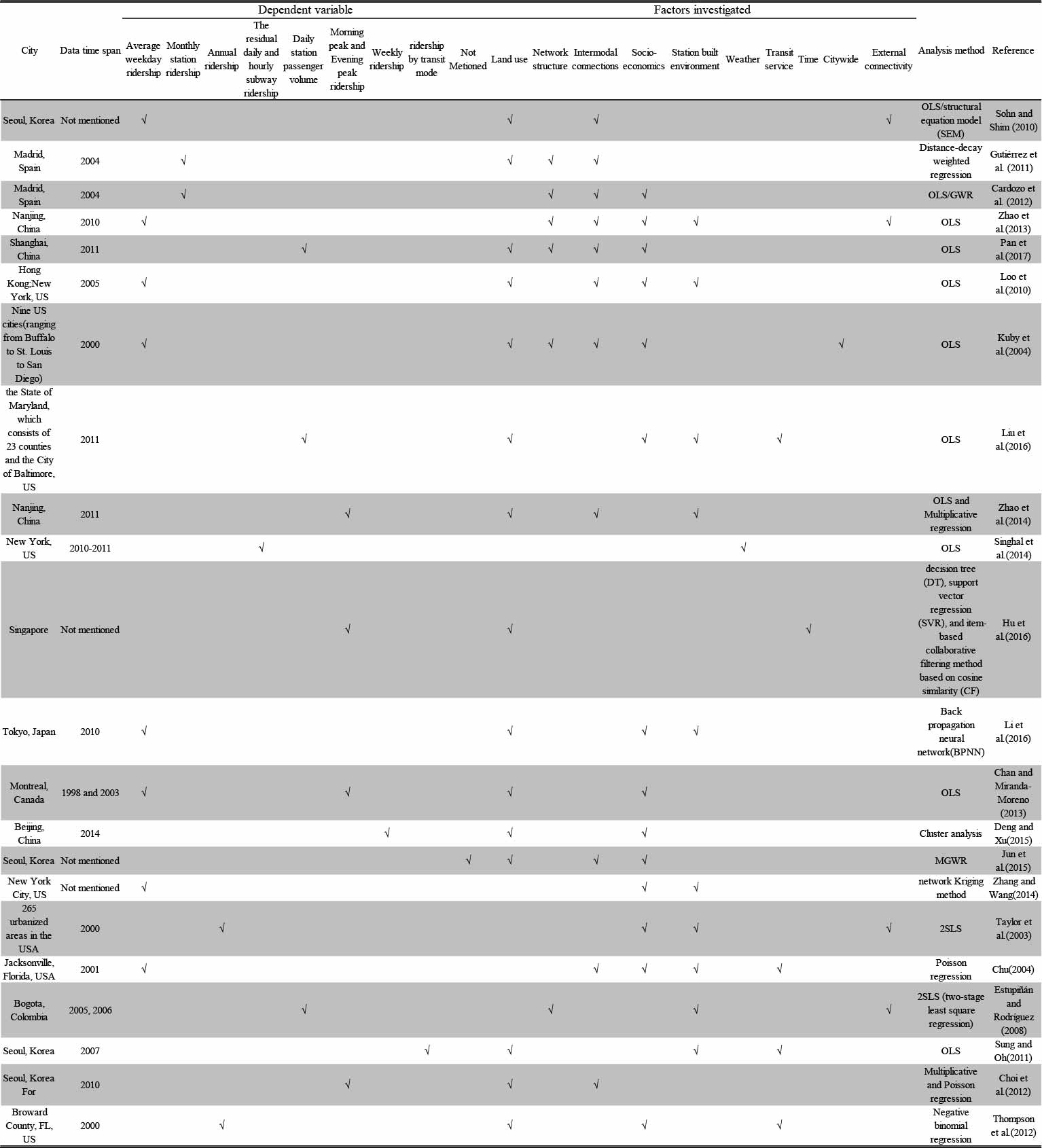}}
	\caption{Summary of literature review on direct demand models for ridership estimation \citep{cardozo2012application,gutierrez2011transit,choi2012analysis,cervero2006alternative,chu2004ridership,walters2003forecasting,kuby2004factors,sohn2010factors,loo2010rail,sung2011transit,thompson2012really,guerra2012half,zhao2013influences,chan2013station,singhal2014impact,liu2016increase,pan2017determines,jun2015land,zhang2014transit,taylor2003analyzing,estupinan2008relationship,deng2015characteristics,li2016forecasting,hu2016impacts}.}
	\label{Fig1}
\end{figure}

%%%%%%%%%%%%%%%%%%%%%%%%
%%%%%%%%%%%%%%%%%%%%%%%%
\bibliographystyle{IEEEtran}
\bibliography{reference}
\end{document}